\documentclass[aps, prb,twocolumn,superscriptaddress, nofootinbib, showpacs]{revtex4-2}

\usepackage{graphicx}
\usepackage{dcolumn}
\usepackage{dcolumn}
\usepackage{textgreek}
\usepackage{textcomp}
\usepackage{adjustbox}
\usepackage{multirow}
\usepackage{mathtools}

\def\KNAO{KNiAsO$_4$}

\def\Tn{$T_N$}

\def\degrees{$^{\circ}$}
\def\iA{\text{\AA}\textsuperscript{-1}}

\def\Rmt{$R\overline{3}$}

\begin{document}

\onecolumngrid
Notice of Copyright This manuscript has been authored by UT-Battelle, LLC under Contract No. DE-AC05-00OR22725 with the U.S. Department of Energy. The United States Government retains and the publisher, by accepting the article for publication, acknowledges that the United States Government retains a non-exclusive, paid-up, irrevocable, world-wide license to publish or reproduce the published form of this manuscript, or allow others to do so, for United States Government purposes. The Department of Energy will provide public access to these results of federally sponsored research in accordance with the DOE Public Access Plan (http://energy.gov/downloads/doe-public-access-plan).
\twocolumngrid

\clearpage

\preprint{APS/123-QED}

\title{Zig-Zag magnetic order and potential Kitaev interactions in the spin-1 honeycomb lattice \KNAO }

\author{K.M. Taddei}
\email{taddeikm@ornl.gov}
\affiliation{Neutron Scattering Division, Oak Ridge National Laboratory, Oak Ridge, TN 37831}
\author{V.O. Garlea}
\affiliation{Neutron Scattering Division, Oak Ridge National Laboratory, Oak Ridge, TN 37831}
\author{A.M. Samarakoon}
\affiliation{Materials Science Division, Argonne National Laboratory, Lemont, IL, 60439}
\author{L.D. Sanjeewa}
\email[corresponding author ]{sanjeewal@missouri.edu}
\affiliation{University of Missouri Research Reactor, University of Missouri, Columbia, MO, 65211}
\affiliation{Department of Chemistry, University of Missouri, Columbia, MO, 65211}
\affiliation{Materials Science and Technology Division, Oak Ridge National Laboratory, Oak Ridge, TN, 37831}
\author{J. Xing}
\affiliation{Materials Science and Technology Division, Oak Ridge National Laboratory, Oak Ridge, TN, 37831}
\author{T.W. Heitmann}
\affiliation{University of Missouri Research Reactor, University of Missouri, Columbia, MO, 65211}
\author{C. dela Cruz}
\affiliation{Neutron Scattering Division, Oak Ridge National Laboratory, Oak Ridge, TN 37831}
\author{A.S. Sefat}
\affiliation{Materials Science and Technology Division, Oak Ridge National Laboratory, Oak Ridge, TN, 37831}
\author{D. Parker}
\affiliation{Materials Science and Technology Division, Oak Ridge National Laboratory, Oak Ridge, TN, 37831}

\date{\today}

\begin{abstract}

Despite the exciting implications of the Kitaev spin-Hamiltonian, finding and confirming the quantum spin liquid state has proven incredibly difficult. Recently the applicability of the model has been expanded through the development of a microscopic description of a spin-1 Kitaev interaction. Here we explore a candidate spin-1 honeycomb system, \KNAO , which meets many of the proposed criteria to generate such an interaction. Bulk measurements reveal an antiferromagnetic transition at $\sim$ 19 K which is generally robust to applied magnetic fields. Neutron diffraction measurements show magnetic order with a $\textbf{k}=(\frac{3}{2},0,0)$ ordering vector which results in the well-known ``zig-zag" magnetic structure thought to be adjacent to the spin-liquid ground state. Field dependent diffraction shows that while the structure is robust, the field can tune the direction of the ordered moment. Inelastic neutron scattering experiments show a well defined gapped spin-wave spectrum with no evidence of the continuum expected for fractionalized excitations. Modeling of the spin waves shows that the extended Kitaev spin-Hamiltonians is generally necessary to model the spectra and reproduce the observed magnetic order. First principles calculations suggest that the substitution of Pd on the Ni sublattice may strengthen the Kitaev interactions while simultaneously weakening the exchange interactions thus pushing \KNAO\ closer to the spin-liquid ground state.

\end{abstract}


\maketitle


\section{\label{sec:intro}Introduction}

The Kitaev Hamiltonian for the spin-$\frac{1}{2}$ honeycomb lattice system is a rare case of an exactly solvable model that also produces exciting physics with imminently useful properties \cite{Kitaev2006}. This confluence of features has driven a massive research effort into studying the phase space and properties of the model and its extensions, searching and designing materials which actually host the prescribed Kitaev interactions, and developing methods to manipulate the resultant topological physics for quantum computation \cite{Savary2016,Chamorro2020,Trebst2022,Broholm2020,Semeghini2021,Takagi2019}. Of this effort, RuCl$_3$ has emerged as a leading candidate material allowing for the observation of many of the proposed experimental properties of the Kitaev quantum spin liquid (QSL) exhibiting fractionalized excitations, fractionalized thermal transport, and leading to the development of extended Kitaev Hamiltonians for real materials which have additional perturbing interactions \cite{Banerjee2016, Banerjee2017, Yokoi2021,Bruin2022}.

Key to this dicussion is the ``Kitaev" interaction which describes an Ising-like interaction along each of the three metal-metal directions for any given site on the honeycomb \cite{Kitaev2006}. This provides an alternate route to a quantum spin-liquid state to the well-known resonating valence state proposed by Anderson as a potential explanation for high temperature superconductivity \cite{Anderson1987}. Such an interaction in a spin-$\frac{1}{2}$ system leads to magnetic frustration and ultimately can give rise to topological classifications, in particular a $Z_2$ classification with bipartite nearly degenerate ground state manifolds classified by their mapping to a torus \cite{Kitaev2006,Wen2002,Chamon2017}. Such a $Z_2$ QSL invokes long-range quantum entanglement which can lead to topological excitations such as the $Z_2$ gauge-flux \lq visons\rq\ and fractionalized excitations including Marjorana Fermions \cite{Chamon2017,Senthil2003,Kitaev2006}. These quasiparticles may exhibit non-abelian statistics which would allow for braiding operations along the particle worldlines and ostensibly a robust form of quantum computation \cite{Kitaev2006,Kitaev2003}.

Yet, despite the promise of these exciting physics, finding candidate materials which exhibit the pure Kitaev interaction has proven challenging \cite{Wen2019,Clark2021,Savary2016,Knolle2019}. Restricting to materials with spin-$\frac{1}{2}$ magnetic ions on honeycomb lattices is quite limiting on its own however, to physically realize the Kitaev interaction while minimizing other interactions an even stricter set of conditions is imposed. Such an optimization requires the magnetic ion be in the $d^5$ valence, be arranged in edge sharing octahedra with perfect 90\degrees\ bond angles, the presence of strong spin-orbit coupling, and a strongly insulating state \cite{Jackeli2009}. Further confounding their realization, the QSL state is both highly sensitive to disorder and often has properties which are trivially emulated by disorder leading to a difficult situation with potential false leads \cite{Rao2021,Huang2021,Zhu2017,Ma2020q}. 

Therefore, expanding the phase space and loosening the restrictions on where one might look to find applications of the Kitaev model and QSL physics is of significant value. Recently, one such effort has been in extending the model to systems with spin ($S$) greater than $\frac{1}{2}$ \cite{Baskaran2008}. However for $S>\frac{1}{2}$, it is not guaranteed that a Kitaev model is realizable, that the resulting state is a QSL, or that such a QSL would host the desired excitations. The key difficulties here are that for $S>\frac{1}{2}$ the Kitaev Hamiltonian is not analytically solvable and until recently no microscopic model for contriving Kitaev interactions for such a state existed. Excitingly, this latter difficultly has been overcome by the development of a microscopic model for a spin-1 system \cite{Stavropoulos2019}. In this model, Kitaev interactions were generated via many of the same requirements as in the spin-$\frac{1}{2}$ case requiring spin-orbit coupling, a honeycomb lattice of edge sharing octahedrally coordinated transition metal ions, and a large Hund's coupling creating a Mott insulator state \cite{Stavropoulos2019}.  Additionally, numerous numerical approaches have been used in attempts to address the solvability issue and have suggested that the $S=1$ case still supports a QSL state indicating a possibility for topological physics and long-range entanglement \cite{Baskaran2008,Khait2021,Dong2020}. Encouragingly, in these studies both ferromagnetic (FM) and anti FM (AFM) Kitaev interactions were suggested to generate a QSL making the spin-1 model relatively permissive \cite{Stavropoulos2019,Dong2020}. 

However, due to the lack of an analytical solution, the nature of the potential QSL state remains contested. Different numerical approaches have led to the emergence of conflicting predictions, with proposals of both gapped and gapless QSL ground states with a possible $Z_2$ topology and topological quasiparticle excitations some of which are not found in the spin-$\frac{1}{2}$ system \cite{Dong2020, Lee2020,Koga2018,Koga2020,Khait2021}. In general, predictions of non-trivial topology seem universal yet despite the expectation for exotic quasiparticles, some analyses have reported that such a QSL may not generate Majorana fermions or non-abelian quasiparticles as may be intuitive considering the spin \cite{Khait2021,Dong2020,Chen2022,Zhu2020,Lee2020}. Nevertheless, it is of great interest to find such a spin-1 Kitaev material to test these predictions and attempt to extend the phase space of QSLs and the general understanding of the Kitaev interaction and Hamiltonian. 

Thus far, few candidate spin-1 honeycomb systems have been found (though there has been success in identifying $S>1$ materials) with Na$_{2-x}$Ni$_2$TeO$_6$ perhaps being the most well studied \cite{Badrtdinov2021,Zhou2021,Chen2021, Samarakoon2021,Bera2022,Lefrancois2016,Kurbakov2020}. This system crystallizes in the $P6_3/mcm$ space group with stacked Ni$^{2+}$ honeycomb layers built of NiO$_6$ octahedra - thus indicating the potential for spin-1 Kitaev physics \cite{Samarakoon2021}. Recently, a comprehensive neutron scattering study on the Na$_2$Ni$_2$TeO$_6$ member found a ``zig-zag" magnetic ground state and a spin-wave spectrum which required the extended Kitaev Hamiltonian to model adding to the material's promise \cite{Samarakoon2021}. However, disagreements between the ground state structure and the energy minimized structure of the spin-Hamiltonian have complicated the analysis and indicated the need for more complex modeling potentially leaving this material further removed from the pure Kitaev state \cite{Samarakoon2021}.

\begin{figure}
	\includegraphics[width=\columnwidth]{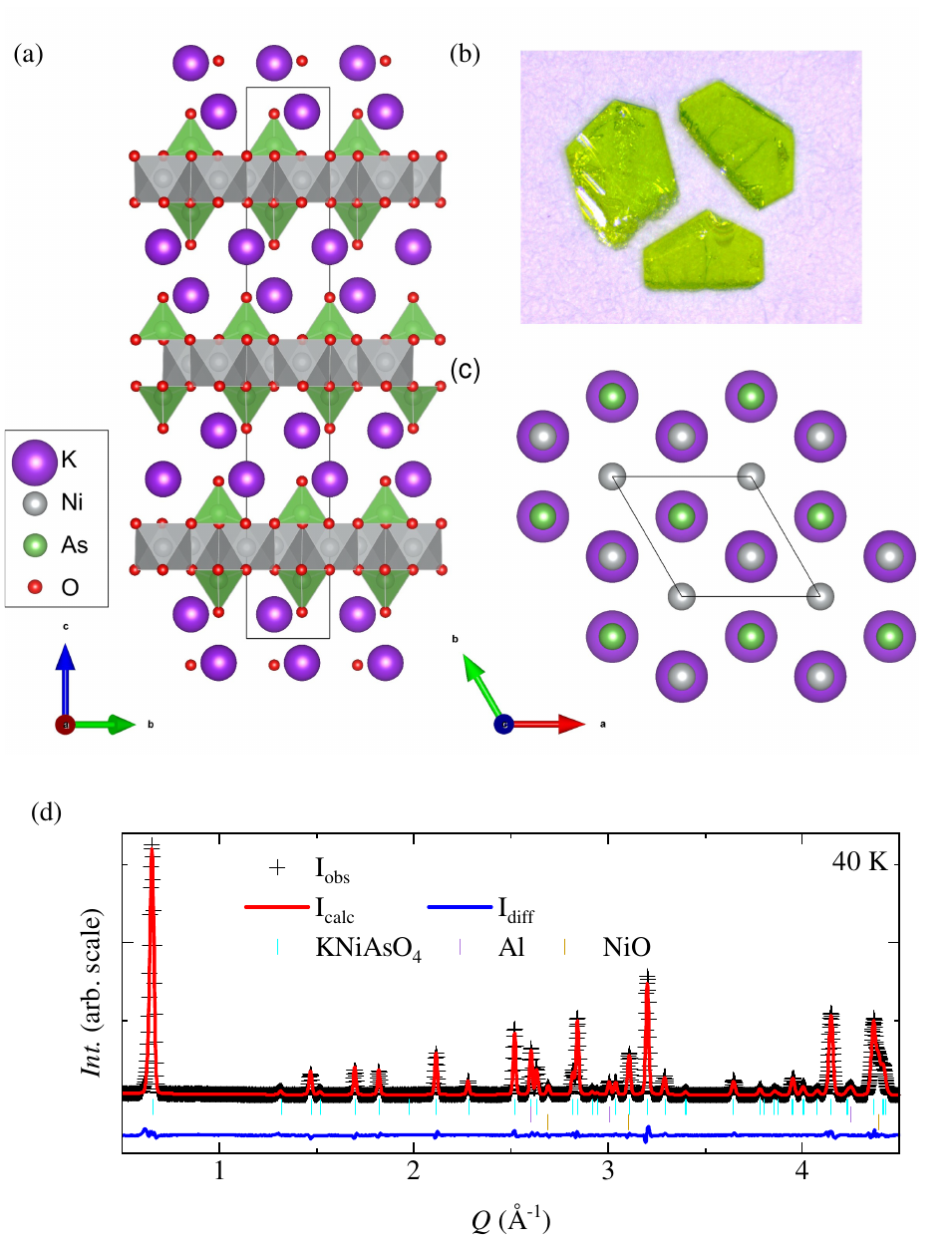}
	\caption{\label{fig:one} (a) Crystal structure of \KNAO\ shown along the $\textbf{a}$ axis illustrating the layered structure. (b) Picture of as grown single crystals of \KNAO\ with the longest edge of the crystals being $\sim 2$ mm .(c) Crystal structure viewed along the $\textbf{c}$ axis. (d) Rietveld profile of the fit to the 40 K neutron powder diffraction data.}	
\end{figure}

Another potential candidate spin-1 honeycomb compound is \KNAO\ which was first discovered and studied several decades ago as a ``mica-like" material with potential for organic ion exchange \cite{Ladwig1979,Buckley1987,Buckley1988,Beneke1982}. \KNAO\ crystallizes in the trigonal $R\overline{3}$ space group (in the hexagonal setting) with a layered structure of edge sharing NiO$_6$ octahedra which create an expected spin-1 Ni$^{2+}$ honeycomb sublattice \cite{Ladwig1979,Buckley1988}. These Ni layers are highly 2D with a large $\sim$ 10 \iA\ interlayer spacing which should strongly suppress interlayer interactions while the overall material has strong insulating characteristics (growing as transparent greenish-yellow crystals) \cite{Ladwig1979}. Thus \KNAO\ appears a good candidate material for exhibiting spin-1 Kitaev physics. 

Unfortunately, early work revealed that \KNAO\ undergoes an AFM transition at $\sim 19$ K thus significantly reducing the chances of finding at QSL state in the as grown compound \cite{Bramwell1988}. Neutron diffraction experiments suggested an ordering vector of $\textbf{k}=(\frac{1}{2},0,0)$ with an AFM chain structure and early theoretical work (absent the Kitaev formulation) even traced out a potential phase diagram based on relative exchange interaction strengths \cite{Bramwell1988}. This early analysis proposed a rich magnetic phase diagram with numerous potential magnetic states in close proximity including the well-known zig-zag structure which is predicted to exist proximate to the QSL state \cite{Bramwell1988}. Therefore, it is worthwhile to reconsider this material to determine if it exhibits any evidence of Kitaev interactions, if so how close to the QSL it might be, and how to optimize the structure and chemistry to tune closer to the pure spin-1 Kitaev model. 

In this paper, we revisit the synthesis, structure, magnetic order, and phase diagram of \KNAO , study the effects of an applied field on the magnetic ground state, and evaluate the potential for Kitaev physics via the experimental determination of the spin-Hamiltonian. We find a ground state magnetic structure of the zig-zag type indicating the potential presence of the Kitaev interaction. Using inelastic neutron scattering we find that the spin-wave spectrum is most consistent with an extended Kitaev Hamiltonian with both Kitaev and Heisenberg interactions. Using first principles calculations, we consider how \KNAO\ might be chemically tuned to strengthen the Kitaev and weaken the Heisenberg interactions. These results suggest \KNAO\ is a useful system to study Kitaev physics in a spin-1 honeycomb lattice.  


\section{\label{sec:methods} Experimental Methods}

\subsection{\label{subsec:synthesis} Synthesis}

Powder samples of \KNAO\ were synthesized using a stoichiometric mixture of KH$_2$AsO$_4$ and Ni(NO$_3$)$_2$6H$_2$O. In a typical reaction, a total of 3 g of KH$_2$AsO$_4$ and Ni(NO$_3$)$_2$ $\cdot$6H$_2$O were mixed in a 1 : 1 stoichiometric molar ratio and ground well using Agate motor and pestle inside an Inflatable Glove Chamber due to both KH$_2$AsO$_4$ and Ni(NO$_3$)$_2\cdot $6H$_2$O being highly hygroscopic. Next, the mixture was loaded in to an alumina crucible and heated to 300 \degrees C overnight. After the initial heating, the mixture was ground and pelletized. These pellets were then re-heated to 900 \degrees C for two days in an Ar-atmosphere after which a homogenous greenish powder was obtained with powder x-ray diffraction confirming the quality and purity of the sample. Once finished, powder samples were transferred to a Ar-filled glove box to prevent moisture absorption. 

Single crystals were synthesized starting with powder samples grown as described above. Using a flux growth, the well ground \KNAO\ powder was mixed with KCl 1:10 mass ratio and sealed in evacuated quartz ampules. The ampoules were then heated up to 900 \degrees C at 60 \degrees C/h and left to dwell at temperature for 7 days. After this the reaction was cooled to 400 \degrees C at 5 \degrees C/hr and then furnace cooled to room temperature. Once at room temperature, high quality green plate-like crystals were recovered from the flux by washing with deionized water. As with the powders, the crystals were stored in the glove box in order to prevent any reaction with air. 

\subsection{\label{subsec:magnetization and heat capacity} Magnetization and Heat Capacity Experiments}

Temperature and field dependent magnetic properties measurements were measured in Quantum Design (QD) Magnetic Properties Measurement System (MPMS). The temperature
dependent static susceptibility ($M/H(T)$) was measured over a temperature range of 2 – 300 K with applied fields between 0.1 and 5 T using single crystal samples with the field oriented parallel and perpendicular to the $\textbf{c}$ axis (i.e. the surface of the plates). Isothermal magnetization measurements were performed at 2 K for fields up to 6 T.
Heat capacity was measured on single crystal samples in the QD Physical Properties Measurement System (PPMS) by the relaxation technique. Measurements were performed in both $\textbf{H}||\textbf{c}$ and $\textbf{H}\perp \textbf{c}$ orientations up to 9 T however, only $\textbf{H}\perp c$ are shown due to the similarity of the signals in either orientation. 

\subsection{\label{subsec:scattering} Neutron Scattering Experiments}

Neutron powder diffraction (NPD) measurements were performed using the constant wavelength powder diffractometer HB-2A of Oak Ridge National Laboratory's (ORNL) High Flux Isotope Reactor (HFIR) \cite{Calder2018}. To prevent grain reorientation and sample movement during measurements in an applied field, the powder sample was pressed into pellets and secured in the sample can with a thin Al post.   Analysis of the neutron powder diffraction data was performed using the Rietveld method as implemented in the FullProf software suite\cite{Rodriguez-Carvajal1993}. For the magnetic structure determination, the Simulated Annealing and Representational Analysis (SARAh) software was used as well as the Bilbao Crystallographic Server \cite{Wills2000, Aroyo2006a,Aroyo2006b,Aroyo2011}. Visualization of the nuclear and magnetic crystal structures was performed using VESTA \cite{Momma2011}. 

Inelastic neutron scattering (INS) experiments were carried out on the HYSPEC direct geometry spectrometer of ORNL's Spallation Neutron Source \cite{Zaliznyak2017}. Measurements were performed on powder samples and the data shown in this work were collected using an incident neutron energy of $E_i$ = 13 meV and a Fermi chopper frequency of 300Hz. This configuration provided an energy resolution of 0.4 meV (determined from the full width at half-maximum at the elastic line). Analysis of the obtained spectra was performed using a machine learning optimization algorithm developed by the authors and linearized spin wave theory as implemented in SPINW \cite{Samarakoon2020,Toth2015}.

\subsection{\label{subsec:firstprinciples} First Principles Calculations}

First principles calculations were performed using the linearized augmented plane-wave (LAPW) density functional theory code WIEN2K, using both the Generalized Gradient Approximation (GGA)  and the commonly used ``GGA+U" approximation, in which an orbital potential $U$ was applied to the magnetic Ni 3$d$ orbitals \cite{Blaha2001,Perdew1996,Anisimov1997}. To avoid, attempting to fit using first principles efforts, only a single $U$ of 5 eV was used in the caluclations. In general, the application of a $U$ further localizes the relevant electrons and hence reduces the exchange energies associated with magnetic order and we will see that this is true here. 

Sufficient numbers of k-points (a minimum of 600 in the relevant Brillouin zone) were used to make the ordered magnetic moments and associated energies of sufficient accuracy - generally to at least 10$^{-2}$ $\mu_B$ and 10$^{-4}$ eV for in-sphere magnetic moment and total energy, respectively. The experimentally determined lattice parameters were used and the internal coordinates relaxed, within the ``straight" GGA approach in an assumed ferromagnetic configuration, until atomic forces were less than 2 mRyd/Bohr. Including magnetic order in the relaxation helps to guard against the risk that the neglect of magneto-elastic effects in non-spin-polarized calculations yields a structure far from the experimental one \cite{Pokharel2018,Niedziela2021, Sanjeewa2020, May2012,Shanavas2014}. Forces in the experimental antiferromagnetic state (see below) remain small. Spin-orbit coupling was not included in these calculations. LAPW Sphere radii of 1.42, 1.57, 1.99 and 2.37 Bohr were used, respectively, for O, As, Ni and K, and an RK$_{max}$ value of 7.0 was employed. This represents the product of the smallest sphere radius - in this case O - and the largest plane-wave expansion wave-vector. Ordinarily for high-precision computational work a larger value (typically 9.0) is more desirable, but in this case the relatively small sphere radius applicable to O means that the effective RK$_{max}$ applicable to the magnetically relevant Ni is in fact well in excess of 9.0, so that one may be reasonably assured of the accuracy of these calculations.

\section{\label{sec:results} Results and Discussion}

\subsection{\label{subsec:rt} Nuclear Structure}

Previous reports suggest \KNAO\ crystallizes in the trigonal \Rmt\ space group (hexagonal setting) in a ``mica-like" layered structure as is shown in Fig.~\ref{fig:one} (a-c) \cite{Bramwell1988,Bramwell1994,Buckley1987,Ladwig1979}. Due to the trigonal symmetry, all four atom species form triangular lattices which are then shifted and stacked along the \textit{c}-axis by the centering operations and mirrored by the rotoinversion. In the case of the Ni site, the \textit{z} position of $\sim$ 0.16 which is close to $\frac{1}{2}$ of $\frac{1}{3}$ leads the layers generated by the centering operations to nearly position on the same plane creating slightly buckled hexagonal layers of Ni. These Ni layers are then sandwiched between O layers describing a hexagonal lattice of NiO$_6$ octahedra. The NiO$_6$ lattice is capped by triangular lattices of As which cover the center of the Ni hexagons making AsO$_4$ tetrahedra (Fig.~\ref{fig:one} (a) and (c)). These Ni-As-O layers stack along the \textit{c}-axis with the interlayers filled by offset triangular lattice K sheets \cite{Bramwell1994,Ladwig1979}. 

To check this structure originally reported several decades ago, we performed NPD at temperatures above the \Tn\ and modeled the resulting pattern using the \Rmt\ structure (as shown in Fig.~\ref{fig:one}(d)). In our Rietveld refinements, the \Rmt\ structure satisfactorily accounted for all observed peak positions and peak intensities save for a small($\sim 1$\%\ volume fraction) impurity phase of NiO and peaks originating from the Al sample can. Therefore, we use this reported structure for the rest of our analysis. The lattice and atomic parameters extracted from the refinements at various temperatures are shown in Table~\ref{tab:one}. Of particular interest here is the sizable \textit{\textbf{c}} lattice parameter which leads to $\sim 10$ \AA\ of separation between the Ni layers and thus should encourage strongly two dimensional physics.

\begin{table}
	\caption{\label{tab:one}Crystallographic parameters of \KNAO at 40 and 2 K under 0 T and at 2 K with a 6 T applied field. The magnetic moment is reported in units $\mu _B/Ni$. }
	\begin{ruledtabular}
		\begin{tabular}{llll}
     		 \multicolumn{1}{c}{ } & \multicolumn{1}{c}{40 K} & \multicolumn{1}{c}{2 K, 0 T} & \multicolumn{1}{c}{2 K, 6 T} \\ 
     		 \multicolumn{1}{c}{ } & \multicolumn{1}{c}{0 T} & \multicolumn{1}{c}{0 T} & \multicolumn{1}{c}{6 T} \\ 
	\hline
	\multicolumn{1}{l}{Space Group} & \multicolumn{1}{c}{\Rmt} & \multicolumn{1}{c}{\Rmt} & \multicolumn{1}{c}{\Rmt} \\
	\multicolumn{1}{l}{$R_{wp}$} & \multicolumn{1}{c}{8.53} & \multicolumn{1}{c}{8.43} & \multicolumn{1}{c}{8.29} \\
	\multicolumn{1}{l}{$\chi^2$} & \multicolumn{1}{c}{17.99} & \multicolumn{1}{c}{16.72} & \multicolumn{1}{c}{3.10} \\
	\multicolumn{1}{l}{$a$ (\AA)} & \multicolumn{1}{c}{4.9864(1)} & \multicolumn{1}{c}{4.9868(1)} & \multicolumn{1}{c}{4.9858(1)} \\
	\multicolumn{1}{l}{$c$ (\AA)} & \multicolumn{1}{c}{28.6182(6)(2)} & \multicolumn{1}{c}{28.6192(6)} & \multicolumn{1}{c}{28.6162(5)} \\
		\multicolumn{1}{l}{$V$(\AA$^3$)} & \multicolumn{1}{c}{616.06(1)} &
	\multicolumn{1}{c}{616.37(1)} & \multicolumn{1}{c}{616.06(1)} \\
	\hline
\multicolumn{1}{l}{K ($6c$)}	&		&		&		\\
\multicolumn{1}{r}{$x$}	&	0	        &      0     	&	0	\\
\multicolumn{1}{r}{$y$}	&	0	        &      0     	&	0	\\
\multicolumn{1}{r}{$z$}	&	0.2893(2)	&	0.2893(2)	&	0.2893(2)	\\
\multicolumn{1}{l}{Ni ($6c$)} 	&		&		&		\\
\multicolumn{1}{r}{$x$}	&	0	        &      0     	&	0	\\
\multicolumn{1}{r}{$y$}	&	0	        &      0     	&	0	\\
\multicolumn{1}{r}{$z$} 	&	0.1653(2)	&	0.1652(1)	&	0.1653(3)	\\
\multicolumn{1}{r}{$m_a$}	&	-	&	-0.2(1)	&	-1.3(1)	\\
\multicolumn{1}{r}{$m_b$}	&	-	&	1.6(1)	&	0.4(1)	\\
\multicolumn{1}{r}{$m_c$}	&	-	&	0.7(1)	&	0.2(1)	\\
\multicolumn{1}{r}{$|\textbf{m}|$} &	-	&	1.8(2)	&	1.5(2)	\\
\multicolumn{1}{l}{As ($6c$)} 	&		&		&		\\
\multicolumn{1}{r}{$x$}	&	0	        &      0     	&	0	\\
\multicolumn{1}{r}{$y$}	&	0	        &      0     	&	0	\\
\multicolumn{1}{r}{$z$}	&	0.5596(1)	&	0.5599(1)	&	0.5596(1)	\\
\multicolumn{1}{l}{O1 ($18f$)} 	&		&		&		\\
\multicolumn{1}{r}{$x$}	&	0.0050(6)	&	0.0060(5)	&	 0.0050(6)	\\
\multicolumn{1}{r}{$y$}	&	0.3447(5)	&	0.3454(4)	&	0.3447(5)	\\
\multicolumn{1}{r}{$z$}	&	0.1255(1)	&	0.1249(1)	&	0.1255(1)	\\
\multicolumn{1}{l}{O2 ($6c$)} 	&		&      	&		\\
\multicolumn{1}{r}{$x$}	&	0	        &      0     	&	0	\\
\multicolumn{1}{r}{$y$}	&	0	        &      0     	&	0	\\
\multicolumn{1}{r}{$z$}	&	0.6178(1)	&	0.6183(1)	&	0.6178(1)	\\

		\end{tabular}
	\end{ruledtabular}
\end{table}

\subsection{\label{subsec:mag} Bulk Magnetic Properties}

\begin{figure}
	\includegraphics[width=\columnwidth]{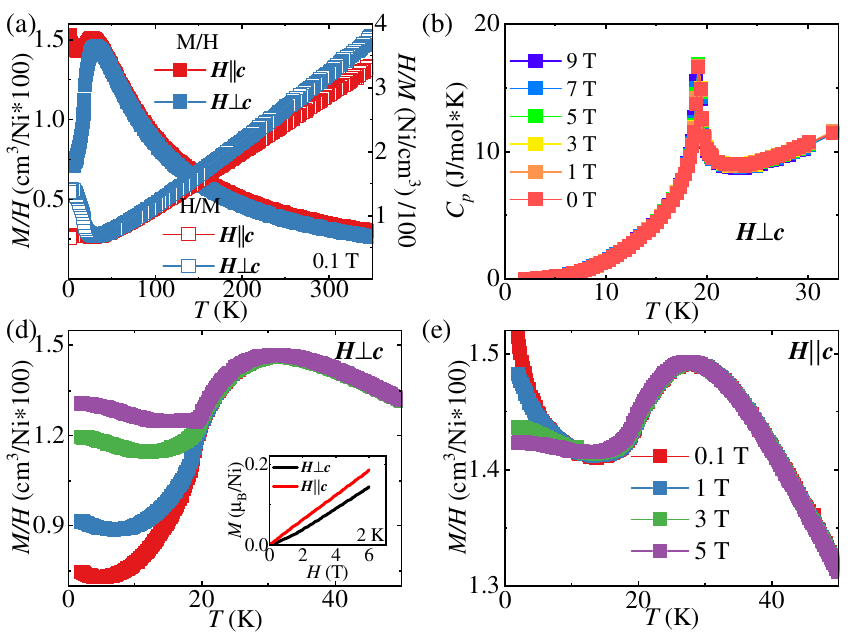}
	\caption{\label{fig:two} (a) Magnetic susceptibility and inverse susceptibility of \KNAO\ collected parallel and perpendicular to the $\textbf{c}$ axis for a probe field of 0.1 T. (b) Heat capacity ($C_p$) measured for $\textbf{H}\perp \textbf{c}$ for applied fields between 0 and 9 T (data collected for $\textbf{H}\parallel \textbf{c}$ is qualitatively similar and so is not shown). Temperature dependent magnetic susceptilibiy collected between 2 and 50 K under applied fields between 0.1 and 5 T for (d) $\textbf{H}\perp \textbf{c}$ and (c) $\textbf{H}\parallel \textbf{c}$. Inset of panel (d) shows the isothermal magnetization collected along the two field orientations. }	
\end{figure}

Previously, \KNAO\ was reported to have an AFM transition at $\sim 19$ K from magnetic susceptibility measurements, here we revisit such measurements with modern instrumentation and expand on them by measuring the field dependence \cite{Bramwell1988}. In Fig.~\ref{fig:two} (a) we show the susceptibility and its inverse as functions of temperature along two different crystallographic directions ($\textbf{H}\parallel \textbf{c}$ and $\textbf{H}\perp \textbf{c}$). We observe a clean AFM-like transition with a broad peak followed by a sharp down turn upon further cooling with a dip at $\sim$ 20 K . Such a broad peak like feature in two-dimensional systems can be indicative of two-dimensional magnetic correlations preceeding the long range three-dimensional order \cite{Hiroi2001,Vasiliev2018}. We can corroborate this interpretation using heat capacity measurements (Fig.~\ref{fig:two}(b)) which show a sharp lambda like anomally at $\sim$ 19 K, indicating the broad peak-like feature is not resultant of the onset of long range order. Comparing the behavior along the different crystallographic directions (i.e. in-plane and along \textit{\textbf{c}}) and find that while the in-plane signal shows a sharp downturn, along-\textit{\textbf{c}} a much smaller drop is observed perhaps indicating a larger moment component in the \textit{\textbf{ab}} plane.

Using the inverse susceptibility curves, we performed Curie-Weiss fitting to extract the effective moment ($\mu_{eff}$) and the Curie-Weiss temperature ($T_{CW}$). We note that in this analysis a high sensitivity of the fit parameters to the fit temperature range was observed (as is common in lower dimensional and frustrated systems), and so a high temperature range of $180 < T < 350$ K was used in the fits \cite{Bramwell1988}. Doing so resulted in fit parameters of $T_{CW} = -0.2 \pm\ 0.5$ K and $\mu_{eff} = 2.7\pm 0.1 \mu_B$ for for fits to the $\textbf{H}\perp \textbf{c}$ inverse susceptibility and of $T_{CW} = -16.1\pm 0.4$ K and $\mu_{eff} = 2.9\pm 0.1 \mu_B$ for the $\textbf{H}\parallel \textbf{c}$ curve. Here the $\mu_{eff}$ is close to that expected for a $S=1$, $3d$ transition metal ion. On the other hand, while both fits produced negative $T_{CW}$, suggesting AFM interactions, the temperatures are quite different. The in-plane $|T_{CW}|$ is two orders of magnitude larger than that of the out-of-plane value, indicating that the magnetic interactions are strongly two-dimensional.  

In Fig.~\ref{fig:two}(d) and (e) we show the low temperature behavior of the transition at different probe fields for the two crystal orientations. For $\textbf{H}\perp \textbf{c}$ we find that increasing field decreases the magnitude of the signal drop after the transition. Similarly, for $\textbf{H}\parallel\ \textbf{c}$ the upturn in the curve at low temperatures decreases with increasing field perhaps indicating that the field is tuning the direction of the moment. However, in either case the \Tn\ shows no significant change even at the highest measured field of 5 T and no evidence of a second transition is observed suggesting only a spin reorientation in the existing magnetic symmetry but no metamagnetic transition \cite{Taddei2022,Taddei2019}. As a further check, the heat capacity was measured as a function of temperature and applied field in both orientations (Fig.~\ref{fig:two}(b) $\textbf{H}\parallel \textbf{c}$ is not shown but looks identical to $\textbf{H} \perp \textbf{c}$) and while a sharp peak is observed at 0 T, it shows no field dependence and no evidence of a second transition is observed at any temperature for fields up to 9 T.

\subsection{\label{subsec:mag} Magnetic Structure and Field Dependence}

\begin{figure}
	\includegraphics[width=\columnwidth]{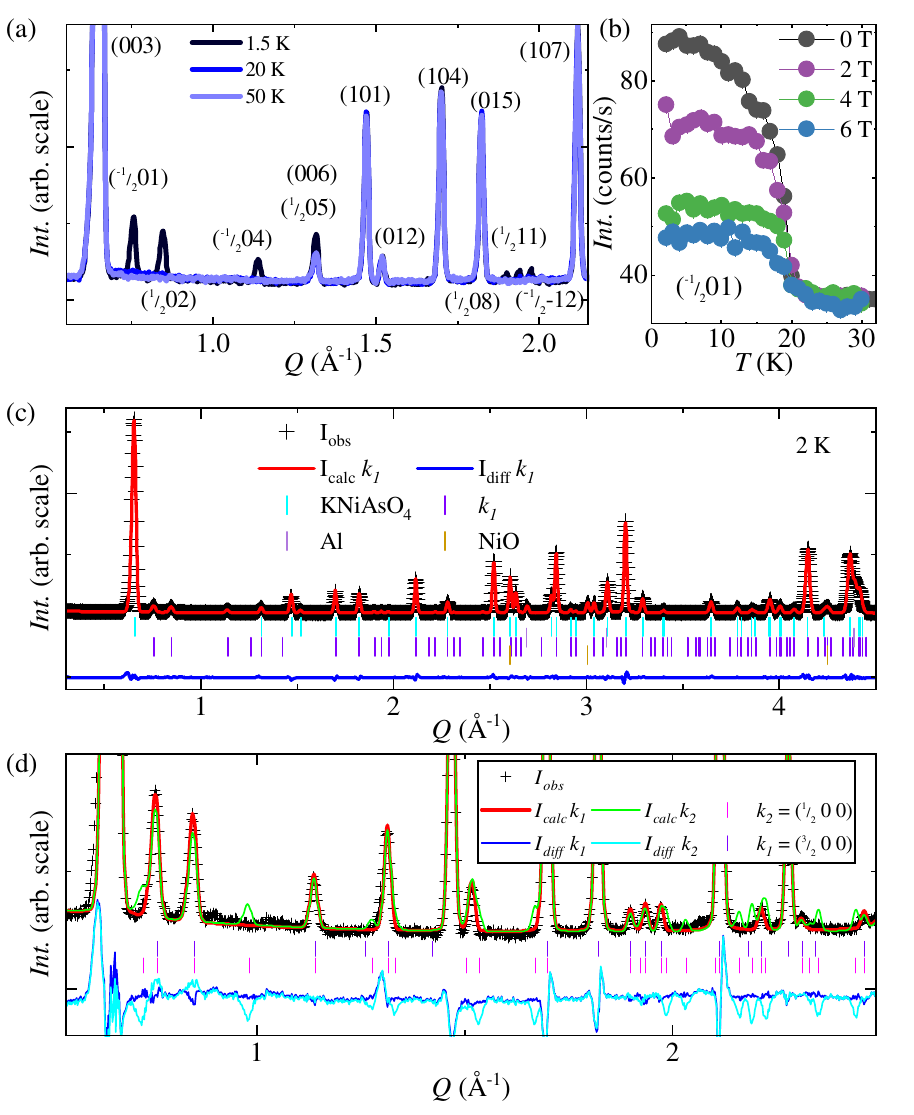}   
	\caption{\label{fig:three}(a) Comparison of powder diffraction patterns collected at 1.5, 20 and 50 K with peak indicies showing the appearance of new fractional integer peaks below \Tn . (b) Temperature dependent order paramete scans for the $(-\frac{1}{2},0,1)$ peak collected on warming under different applied fields. (c) Rietveld profile to the 2 K diffraction pattern for the best fit magnetic model. (d) Rietveld profile comparison of the previously reported magnetic structure (with $\textbf{k}=(\frac{1}{2},0,0)$) and the $\textbf{k}=(\frac{3}{2},0,0)$ structure reported here.}	
\end{figure}

With the bulk measurements indicating an AFM transition at $\sim$ 20 K we next turn to neutron powder diffraction to solve the magnetic structure. In Fig.~\ref{fig:three}(a) we show neutron powder diffraction patterns collected above and below the signal observed in the susceptibility. Upon cooling below \Tn\ numerous new reflections appear which cannot be indexed by the nuclear structure indicating AFM ordering. To better associate these new peaks with the signal seen in the bulk measurements, an order parameter scan was performed by collecting the intensity of the peak at $\sim$ 0.76 \iA\ as a function of temperature (with the intensity of a magnetic Bragg peak being proportional to the square of the ordered moment) as shown in Fig.~\ref{fig:three}(b). The peak intensity is seen to grow below 20 K consistent with the behavior observed in the susceptibility. 

To solve the magnetic structure, we first identify ordering vectors which can account for the observed reflections. As shown in Fig.~\ref{fig:three}(a) all the additional low temperature peaks can be indexed by $\textbf{k}=(\frac{3}{2},0,0)$ (we note that here $\textbf{k}=(\frac{3}{2},0,0)$ is indicated rather than $\textbf{k}=(\frac{1}{2},0,0)$ due to the observed extinction rule $-h+k+l=3n$). Using representational analysis, this $\textbf{k}$ together with the \Rmt\ space group and $6c$ Wyckoff position of the Ni$^{2+}$ ion allows two possible irreducible representations (irreps $\Gamma$) each of which split the $6c$ site into two symmetry related sites (one with the original position and one at $x,y,-z$) with a total of three basis vectors - one each for the three crystallographic directions (see Table~\ref{tab:two}). Of the two $\Gamma$, the first gives rise to a ``stripy" type magnetic order with AFM along the zig-zag chains in the [010] direction of the honeycomb lattice and FM correlations between the chains. The second $\Gamma$ describes a ``zig-zag" order with FM along the chains and AFM between them. Both of these $\Gamma$ correspond to the magnetic space group $P_S\overline{1}$ with a doubling of the \textit{a} lattice parameter and an origin shift of $(\frac{1}{2},0,0)$ in the case of $\Gamma_2$.

\begin{table}
	\caption{\label{tab:two} Irreducible representations ($\Gamma$), basis vectors ($\psi$), magnetic space group (MSG), magnetic super cell, and origin shift for $\textbf{k} = (\frac{3}{2},0,0)$. Basis vectors contain components for the two generated Ni sites, the original at $(0,0,0.165)$ indicated by $\psi_1$ and a second at $(0,0,0.835)$ indicated by $\psi_2$.}
	\begin{ruledtabular}
		\begin{tabular}{ccccccc}
    		 \multicolumn{1}{c}{$\Gamma$} & \multicolumn{1}{c}{$\psi_1$} & \multicolumn{1}{c}{$\psi_2$}  & \multicolumn{1}{c}{MSG} & \multicolumn{1}{c}{Super cell} & \multicolumn{1}{c}{Origin shift} \\
	\hline
	\multirow[t]{3}{*}{$\Gamma_1\ $} & (1,0,0) & (1,0,0)   & \multirow[t]{3}{*}{$P_S\overline{1}$} &   ($2a,b,c$) & (0,0,0) \\
									 & (0,1,0) & (0,1,0)   &          &   & \\
									 & (0,0,1) & (0,0,1)   &          &   &  \\
	\multirow[t]{3}{*}{$\Gamma_2\ $} & (1,0,0) & (-1,0,0) & \multirow[t]{3}{*}{$P_S\overline{1}$} & ($2a,b,c$)  &  ($\frac{1}{2}$,0,0) \\
									 & (0,1,0) & (0,-1,0)   &          &   &  \\
									 & (0,0,1) & (0,0,-1)  &          &   &  \\
		\end{tabular}
	\end{ruledtabular}
\end{table}

\begin{figure}
	\includegraphics[width=\columnwidth]{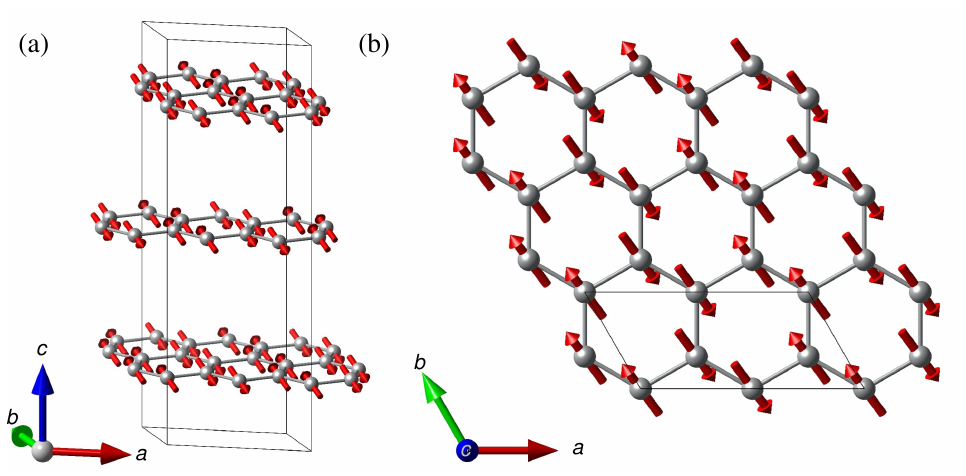}
	\caption{\label{fig:four} Best fit magnetic structure of \KNAO\ oriented to visualize the (a) stacking relationship of the Ni planes along the $\textbf{c}$ direction and (b) the magnetic order in a single Ni layer.}	
\end{figure}

Rietveld refinements were performed using both of these models and the zig-zag structure of $\Gamma_2$ was found to produce a decisively better fit to the experimental pattern. The resulting profile together with the data are shown in Fig.~\ref{fig:three}(c) and the extracted crystallographic and magnetic structure parameters are reported in Table~\ref{tab:one}. The obtained magnetic structure (Fig.~\ref{fig:four}), has FM zig-zag chains running along the [010] direction of the unit cell with AFM coupling to the neighboring chains (i.e. along $\textbf{a}$). The $P_S\overline{1}$ MSG allows the magnetic moments to have non-zero components in all three crystallographic directions and in our refinements we found it necessary to allow all three components to be non-zero to obtain the best fit. In the refined structure, the majority of the moment lies in the \textit{\textbf{ab}}-plane with a small component along \textit{\textbf{c}} describing a $\sim 22$\degrees\ canting (see Table~\ref{tab:two}). The magnitude of the ordered moment is $1.8(2) \mu_B/$Ni which is smaller than the effective moment but very close to the expected size for a spin-1 system \cite{Bramwell1994,Bramwell1988}. 

We note that the magnetic structure found here is different than that reported in ref.~\onlinecite{Bramwell1994} and so it is worth considering this previous model more carefully. In ref.~\onlinecite{Bramwell1994} a $\textbf{k}=(\frac{1}{2},0,0)$ was suggested with a magnetic structure that doubled the cell along \textit{\textbf{a}} (or \textit{\textbf{b}}) and recreated a stripy-type structure. In our analysis using representational analysis, $\textbf{k}=(\frac{1}{2},0,0)$ breaks the $6c$ Wyckoff site into two separate orbits (or symmetry distinct atomic sites) with the same positions as described for the symmetry related atoms in the $\textbf{k}=(\frac{3}{2},0,0)$ structures and produces only a single $\Gamma$ for either orbit. This $\Gamma$ has three basis vectors per site - one for each crystallographic direction leading to a total of six parameters and allowing the two sites to have differently sized ordered moments. Using this model we again performed Rietveld refinements and the best fit is shown in Fig.~\ref{fig:three}(d) together with the fit from the $\textbf{k}=(\frac{3}{2},0,0)$. Here it is clear that the $\textbf{k}=(\frac{3}{2},0,0)$ zig-zag model accounts for both the observed peak positions and intensities better than the $\textbf{k}=(\frac{1}{2},0,0)$ model with the latter adding magnetic peaks where no new intensity is seen while also not correctly fitting the observed intensities. We suggest that this discrepancy with the previous report likely results from our use of a higher intensity, higher resolution instrument which allowed us to observe many more magnetic peaks than in the previous work as well as the development of more powerful tools for magnetic structure solution in the intervening time. Therefore, we continue our analysis using our zig-zag magnetic structure. 

\begin{figure}
	\includegraphics[width=\columnwidth]{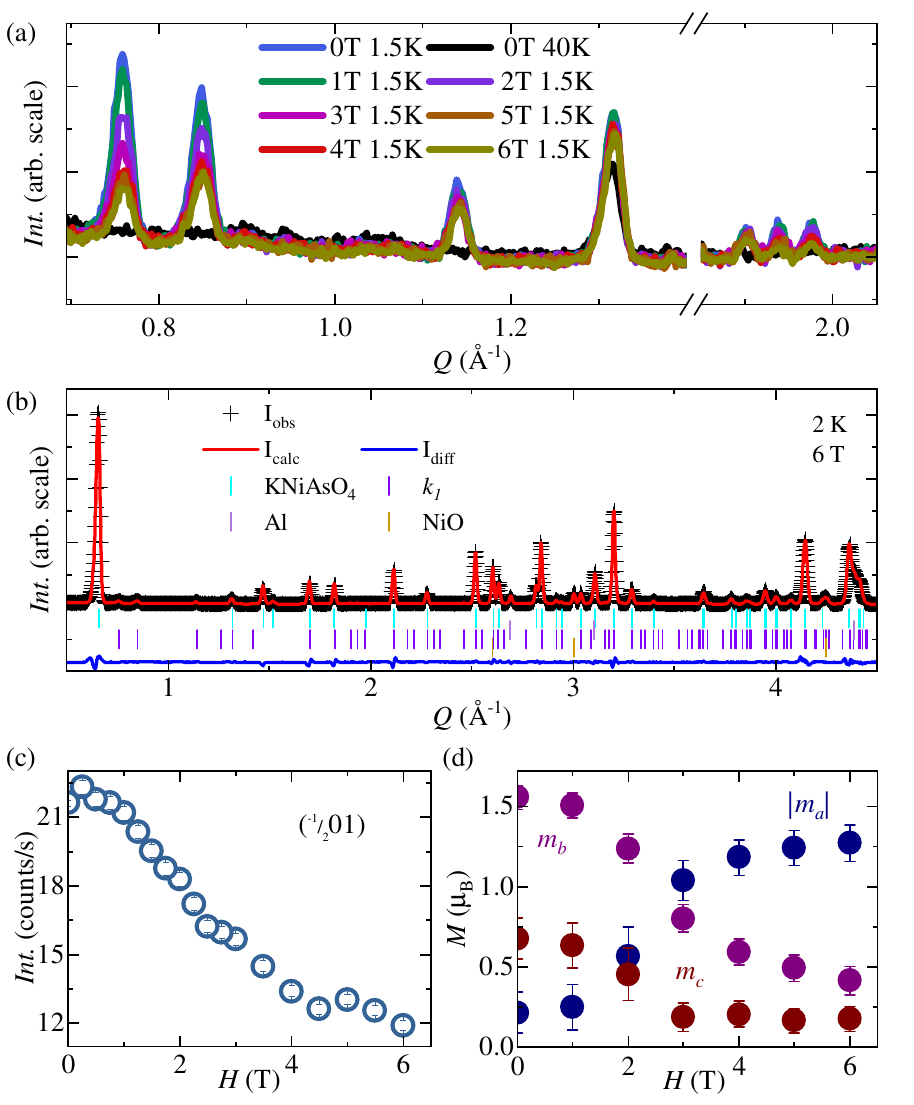}
	\caption{\label{fig:five} (a) Powder diffraction patterns under different applied fields. (b) Rietveld profile fit of data collected at 2 K under a 6 T applied field using the magnetic structure determined from the analysis of the 0 T data. (c) Order parameter scan collected upon warming from 2 K on the $(-\frac{1}{2},0,1)$ magnetic peak. (d) Refined components of the Ni$^{2+}$ magnetic moment as a function of applied field.}	
\end{figure}

Having established the zig-zag structure as the zero-field magnetic ground state, we turn to the field dependent structure. Shown in Fig.~\ref{fig:five} (a) are several powder patterns collected under different applied fields together with the 0 T, 1.5 K and 0 T, 40 K data for comparison. Focusing on the most intense low angle peaks (the $(-\frac{1}{2},0, 1)$ and $(\frac{1}{2}, 0, 2)$ reflections) we see that as the field is increased the magnetic intensity monotonically decreases to the highest measured field of 6 T (Fig.~\ref{fig:five}(c)). However, the reduction in intensity is not identical between the peaks with the $(-\frac{1}{2}, 0, 1)$ changing from the more intense peak at 0 T to the weaker reflection by 6 T. Similarly, considering the three reflections between 1.9 and 2 \iA\ we see the intensity of the two higher angle peaks decrease with increasing field while the lower angle peak remains nearly constant. These observations imply that the increasing field has two effects on the magnetic structure, reducing the ordered moment and changing its direction with respect to the nuclear structure, both of which are broadly consistent with the magnetization data previously discussed. This is further supported by considering the temperature dependence of the $(-\frac{1}{2}, 0,1)$ peak under different applied fields (Fig.~\ref{fig:three}(b)) and at 2 K as a function of field (Fig.~\ref{fig:four}(c)). In the former, as in the heat capacity measurements, we see no change in the transition temperature. In the latter we see a continuous decrease in intensity to the highest measured field. In either case, no sign of a phase transition is observed.  

Ideally, one would like to extract the behavior of the magnetic field along different crystallographic directions however, in this work such measurements were not feasible. Therefore, despite the orientational averaging of the powder data, we nevertheless performed quantitative analysis by modeling the magnetic scattering observed in the data collected under an applied field. We note that since no metamagnetic transition was observed, which in such measurements can lead to multiple macroscopically coexisting magnetic structures due to the different field orientations of different powder grains (see for instance ref.~\onlinecite{Xing2021}), analysis of the field dependent data may just give additional insights to which crystallographic directions are more magnetically soft.      

In Fig.~\ref{fig:five}(b) we show a fit using the zig-zag model of the data collected at 1.5 K and 6 T. As seen, despite the discussed difficulties, the calculated intensities match the experimental profile remarkably well without the need for additional magnetic phases to account for different relative field orientations, with similar quality fits obtained for all measured fields. To elucidate the above described field dependence, we extracted the independent components of the refined magnetic moment (Fig.~\ref{fig:five}(d)). As the field is increased, we find a monotonic decrease in the magnitude of both $m_b$ and $m_c$ while the magnitude of $m_a$ increases thus describing a rotation of the moments to be more in-plane and more perpendicular to the chain direction. We note that in our refinements we also found the field to drive a slight contraction of both the \textit{a} and \textit{c} lattice parameters which both continuously decrease with increasing field.

Finding the zig-zag order in \KNAO\ is quite interesting. In both the spin-1/2 and spin-1 Kitaev models on a honeycomb lattice, the zig-zag magnetic ground state is expected to be adjacent to the Kitaev state \cite{Stavrou2015,Chaloupka2013,Dong2020}. In calculations, the zig-zag structure is found to stabilize from a spin-Hamiltonian which contains both Kitaev and Heisenberg interactions \cite{Stavrou2015}. Thus, perhaps the observed structure is evidence of a proximity to Kitaev physics in \KNAO , a case we consider more rigourously in the next section .

\subsection{\label{subsec:INS} Spin-waves and the Spin-Hamiltonian}

\begin{figure}
	\includegraphics[width=\columnwidth]{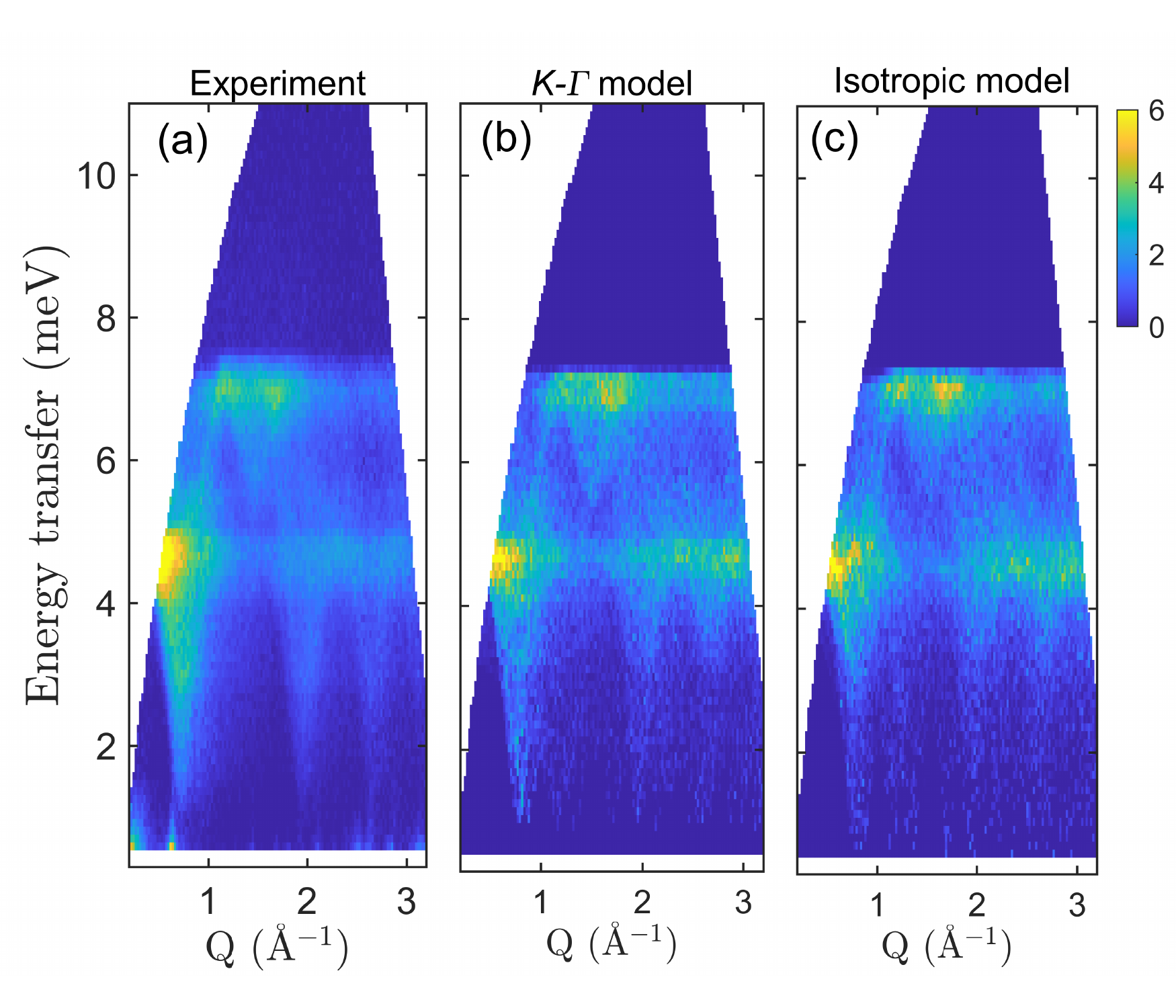}
	\caption{\label{fig:six} (a) Powder inelastic neutron spectrum ($S(q,\Delta E)$) of \KNAO\ measured on HYSPEC using $E_i=15$ meV at 2 K. Calculated INS spectra from best fit parameters for the (b) $\Gamma - K$ and (c) ISO spin-Hamiltonians.}	
\end{figure}

While the realized long range order allows us to place \KNAO\ on established generic phase diagrams of the perturbed Kitaev Hamiltonian, to better elucidate whether it actually exhibits Kitaev interactions it is necessary to study the spin-dynamics of the system. Shown in Fig.~\ref{fig:six}(a) is the INS spectrum for a powder sample of \KNAO\ collected at 2 K using the HYSPEC spectrometer with an incident energy of $E_i = 13$ meV. Here, we see well defined excitations with several acoustic modes arising from $\sim 0.75, 2,$ and $2.6$ \iA\ and optical modes around 4.5 and 7 meV as expected for an AFM structure. At low energies the acoustic modes are seen to have an energy gap of $\sim 1$ meV indicating some explicit symmetry breaking in the underlying spin-Hamiltonian. 

Such behavior is unlike that of the canonical proximate QSL $\alpha$-RuCl$_3$ where even powder INS evidenced heavily damped, broad excitations indicative of the QSL state, though it is worth noting that in the spin-1 QSL such fractionalized excitations may not be expected in the spin-channel \cite{Banerjee2016,Chen2022}. Rather here the spectrum looks well ordered as seen recently in other Kitaev candidate materials such as Na$_2$Co$_2$TeO$_6$, Na$_2$Ni$_2$TeO$_6$, BaCo$_2$(AsO$_4$)$_2$ and Na$_2$IrO$_3$ \cite{Samarakoon2021,Choi2012,Halloran2022}. Nevertheless, it is important to determine whether the found ordered state results from Kitaev interactions to understand if and how one could tune the interactions towards a Kitaev state.

To explicate the spin-Hamiltonian, linear-spin-wave theory (as implemented in SPINW) was used to fit the observed spectra with several different trial Hamiltonians. Of these, here we focus on the extended Kitaev ($\Gamma - K$) and the isotropic (ISO) Heisenberg models as they provided the best fits of the observed spectra (we note that an \textit{XXZ}-model produced similar quality fits to the ISO model but with less reasonable parameters) \cite{Jackeli2009, Samarakoon2021,Rau2014}. A generalized form of the spin-Hamiltonaian is shown in Eq.~\ref{eq:Ham}.  

\begin{equation}
	\label{eq:Ham}
	\begin{multlined}
	\mathcal{H}_{\Gamma-K} = \sum_{\langle ij\rangle} [J \vec{S_i}\cdot \vec{S_j} + K S_i^\gamma S_j^\gamma \\ + \Gamma (S_i^\alpha S_j^\beta +S_i^\beta S_j^\alpha) \\ + 
	\Gamma ' (S_i^\alpha S_j^\gamma +S_i^\gamma S_j^\alpha +S_i^\beta S_j^\gamma + S_i^\gamma S_j^\beta)] \\  - 
	D \sum_{i} (S_i \cdot \tilde{n}_i)^2
	\end{multlined}
\end{equation}    

\begin{figure}
	\includegraphics[width=\columnwidth]{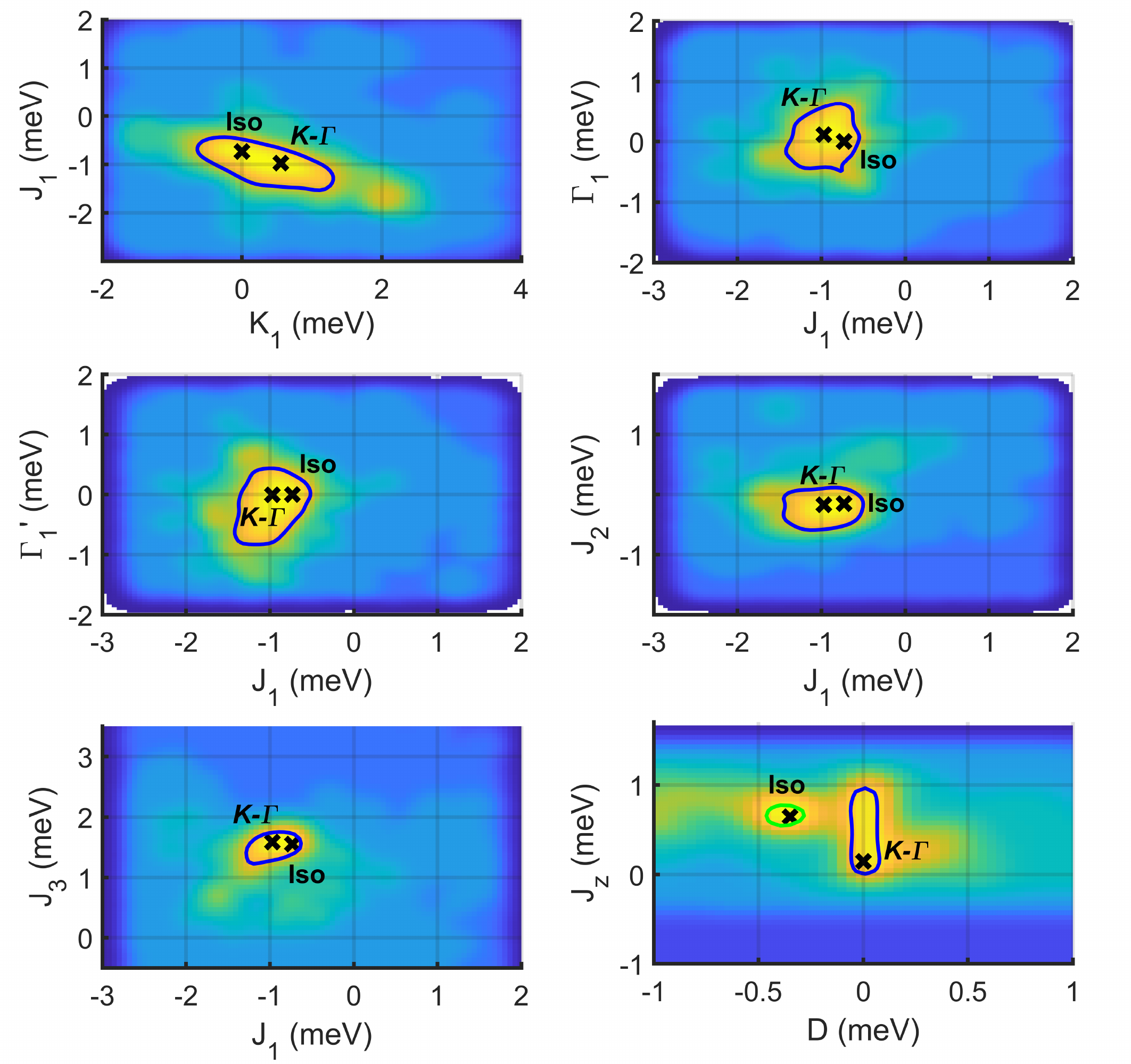}
	\caption{\label{fig:seven} Two dimensional cuts of the higher dimensional spin-Hamiltonian optimization manifold with the color scale indicating the $\chi ^2$ of the given parameter's modeling of the observed spectra. Small and large values of $\hat{\chi} ^2_{INS}$ are indicated by yellow and blue respectively on a logarithmic scale.}	
\end{figure}

In Eq.~\ref{eq:Ham}, the indicies $<ij>$ run over first, second, third in-plane nearest neighbor (NN) interactions and a single inter-layer interaction with associated exchange interactions of $J_1, J_2,J_3$ and $J_z$ respectively. The $\alpha , \beta$ and $\gamma$ label the three Kitaev spin directions and the $K, \Gamma , \Gamma ' , D$ and $\tilde{n}_i$ denote the Kitaev, symmetric off-diagonal exchange, asymmetric off-diagonal exchange, single-ion anisotropy (SIA), and the SIA direction, respectively. The $D$ term is added for generality and is only non-zero when $K \rightarrow 0$ where $H_{\Gamma - K}$ reduces to the Heisenberg \textit{XXZ}-like model which can be further reduced to the ISO model by setting the off-diagonal terms ($\Gamma$ and $\Gamma'$) to zero. To apply such a highly parameterized model (with a maximum of seven parameters) we used an iterative machine-learning optimization procedure together with SPINW as explained in ref.~\onlinecite{Samarakoon2020} and further discussed in ref.~\onlinecite{Samarakoon2021}. Fig.~\ref{fig:seven} shows select slices of the resulting manifold of possible solutions plotted as contours of the cost-estimator $\hat{\chi} ^2 _{INS}$ function (which is determined from the cost-function $\chi ^2 = \sum_{\omega}\sum_{Q}(I_{exp}(Q,\omega)-I_{calc}(Q,\omega))^2$ using the process described in ref.~\onlinecite{Samarakoon2021}). Using this cost-estimator, the best solutions modeling the observed INS spectra were chosen at $\hat{\chi} ^2 _{INS}$ minima (indicated by the yellow regions in Fig.~\ref{fig:seven}) and are labeled as ISO and $\Gamma -K$.        

Fig.~\ref{fig:six}(b) and (c) shows the resulting simulated INS spectra calculated from the best fit models of both the extended Kitaev and the ISO Hamiltonians (i.e. $D=0$ and $K,\Gamma,\Gamma'=0$ respectively) with the refined parameters shown in Table~\ref{tab:two}. Surprisingly, both models are able to reproduce fairly well most of the observed scattering despite having very different physical origins - possibly due to the information lost during the powder averaging of the INS spectra. Both models reproduce the flat optical modes at $\sim 7.5$ and $\sim 4.5$ meV and capture the observed dispersions for the acoustic modes. Similarly, both models recreate the observed peak intensities near $(q,\Delta E) = (0.5$\iA $, 4.5$ meV$)$ and in the high energy optic mode. However, at lower energies, $\Gamma - K$ is better able to account for the observed gap at $q\sim 0.8$\iA , though we note that both models produce gaps in this range. More significantly, the ISO model requires a SIA which either under estimates or grossly overestimates the gap and is always found to produce a magnetic structure with Ni$^{2+}$ moments along the \textit{\textbf{c}}-axis even in solutions with easy-plane SIA. Such features are manifestly inconsistent with the structure found in the NPD analysis. On the other hand, the extended Kitaev model naturally allows for the observed canted structure, thus strongly indicating its applicability for this system.

\begin{table}
	\caption{\label{tab:two}Parameters obtained from the extended Kitaev Hamiltonian optimized to reproduce the observed spin-wave spectrums. For the $\Gamma - K$ model the $D$ term is fixed to zero while for the ISO model the $K$ and $\Gamma$ terms are fixed to zero. In the $K-\Gamma$ model the $\Gamma '$ term was found unnecessary and so set to zero in the final optimization. All values are in meV.}
	\begin{ruledtabular}
		\begin{tabular}{cdddddddd}
    		 \multicolumn{1}{c}{Model} & \multicolumn{1}{c}{$J_1$}  & \multicolumn{1}{c}{$K_1$} & \multicolumn{1}{c}{$\Gamma_1$} & \multicolumn{1}{c}{$\Gamma_1'$} & \multicolumn{1}{c}{$J_2$} & \multicolumn{1}{c}{$J_3$} & \multicolumn{1}{c}{$J_z$} & \multicolumn{1}{c}{$D$}\\
	\hline
	 $K-\Gamma$ & -0.97 & 0.56 & 0.11 & 0 & -0.17 & 1.6  & 0.15 & 0     \\
	 $ISO$      & -0.73 & 0    & 0    & 0 & -0.15 & 1.55 & 0.65 & -0.35 \\

		\end{tabular}
	\end{ruledtabular}
\end{table}

Considering the obtained parameters we find that the Kitaev model produces AFM Kitaev interactions with FM $J_1$ and $J_2$ exchange interactions and AFM $J_3$ and inter-layer $J_z$ interactions. Here $J_1$ and $J_3$ interactions are dominant with the Kitaev term being roughly half that of the $J_1$. This indicates that even in the extended Kitaev model the Kitaev term is not the leading interaction unlike other candidate compounds \cite{Samarakoon2021,Banerjee2016}. For the ISO model, we obtain qualitatively similar values for the exchange interactions albeit with a large increase in the inter-layer interaction $J_z$ which surprisingly becomes similar in size to the $J_1$ exchange. 

Interestingly, the zig-zag structure found in our diffraction analysis was predicted to arise in a spin-1 extended Kitaev model for $J/K < 0$ \cite{Stavropoulos2019,Khait2021}. This is consistent with the $J/K$ determined in our spin-wave fit using the $\Gamma -K$ Hamiltonian further indicating consistency to this model. For the ISO-model, the large $J_{z}/J_{1,2,3}$ ratio is unexpected. For similar layered transition metal honeycomb lattices the inter-layer exchange interactions are typically an order of magnitude smaller than the NN exchange \cite{Matsuda2019, Samarakoon2021,Halloran2022,Gao2021,Choi2012}. For \KNAO , the inter Ni layer distance is $\sim 10$ \iA\ and a super exchange pathway would entail electron exchange through both the As and K layers, that such a large $J_z$ would result is surprising and unlikely lending support to the extended Kitaev model. Considering the slightly better visual agreement of the $\Gamma -K$ model, its consistency with the observed magnetic structure, and its more physically reasonable exchange parameters we tentatively suggest it as the more appropriate model for this system. However, additional INS studies using single crystal samples are needed to more firmly corroborate this interpretation.

\subsection{\label{subsec:DFT} First Principles Analysis of the Magnetic Structures}

First principles calulations were carried out to further elucidate the microscopic origins of the zig-zag magnetic structure. Using DFT, three magnetic states were studied - a non-spin-polarized state (NM), the FM state and a nearest neighbor antiferromagnetic state (AF1) which approximates the neutron-observed ground state. We have not attempted to estimate the interlayer coupling but expect it to be quite small in view of the large Ni-Ni distances involved.  However, we note that in recent work (refs.~\onlinecite{Chen2019,Yan2020}) even nearly negligible interlayer couplings were found sufficient to stabilize magnetic order with appreciable ordering points of order 10 K or higher, as is possibly the case here.   

As expected for an oxide, even within the straight GGA the magnetic order is local in nature and falls well below the non-magnetic state NM in energy, with the FM state falling, within GGA,  over 800 meV/Ni below the NM state and the AF1 ground state state some 12.9 meV/Ni below the FM state indicating the robustness of the AF1 state. Within the GGA+U this latter energy difference is reduced to some 3.3 meV/Ni. Consistent with the expected +2 charge state and the 3$d^8$ electronic configuration of Ni we find a moment of 2 $\mu_B$ per Ni, i.e the total cell moment is 2.00 $\mu_B$ for the FM state, with the in-sphere moment, as expected, being slightly smaller at 1.59 $\mu_B$/Ni in GGA and 1.74 in GGA+U. The corresponding moments for the AF1 state are 1.58 (GGA) and 1.74, the (GGA+U) relative insensitivity of in-sphere magnetic moment to magnetic configuration characteristic of local moment magnetism and in good agreement with the value determined from the Rietveld modeling (of 1.8(1) $\mu_B$).  The AF1 ground state is insulating, with band gaps of 1.71 eV within GGA and 3.07 within GGA+U.

The observed energetics would indicate estimated magnetic ordering points between 12.6 K (GGA+U) and 50 K (GGA), with the observed ordering point of 19 K bracketed by these values \cite{Williams2016, May2012, Lamichhane2016, Klepov2021}. That the lower GGA+U value is significantly closer to the experimental value is a fair suggestion that the GGA+U-depicted localization of Ni orbitals is active here. Direct linear interpolation of the band-gap values based on the N\'eel point prediction suggests that the experimental band-gap is likely close to 3 eV. 

Next we consider via first principles how to push \KNAO\ closer to the Kitaev model. In this model the previously discused ratio $J/K$ plays a key role. In the true Kitaev limit $J/K$ approaches zero, and so in an attempt to discover Kitaev physics here we have briefly explored substitutions designed to increase $K$ and potentially decrease $J$. For this purpose we have substituted, in separate calculations, one of the 2 in-cell Ni by the isoelectronic Pd and one of the As by the similarly isoelectronic Sb. As heavier elements, both substitutions would be expected to increase spin-orbit coupling and thus potentially the $K$ interaction, and the generally weaker magnetic coupling of elements such as Pd would be expected to also decrease J. For Sb alloying, the calculation immediately crashes, suggestive that the much greater volume of Sb may impede its substitution for As. For the Pd case, however, something interesting happens. Despite the calculation's initialization with the Pd moment oriented opposite to Ni, the Pd moment rapidly \lq flips\rq\ to become ferromagnetically coupled to the Ni, and converges to a ferromagnetic solution with the same moment (4 $\mu_B$/cell) as in the pure-Ni FM case. The former fact means that Pd substitution can be expected to move the system towards itinerant behavior, with all the potential complexities associated with itinerant physics.  The latter, corresponding point means that Pd substitution may drive the systems towards quantum-critical behavior by frustrating the normally antiferromagnetic Ni-Ni nearest-neighbor interaction and thereby driving the ordering point towards T=0. We leave this observation for future experimental efforts to exploit.

\section{\label{sec:con} Conclusions}

We report the synthesis and comprehensive magnetic analysis of the spin-1 honeycomb compound \KNAO . Crystallizing with the $R\overline{3}$ space group symmetry \KNAO\ presents a nearly perfect layered honeycomb material with nearly planar Ni honeycomb and a large $\sim 10 $ \iA\ interlayer spacing. Charge counting suggests a Ni$^{2+}$ state with a $5d^8$ valence giving rise to a spin-1 system. The octahedral NiO$_6$ enviornment has nearly perfect 90\degrees\ Ni-O-Ni bond angles which together with the other chemical and structural components set the stage to realize the highly desired Kitaev interaction. Magnetization and heat capacity measurements reveal an AFM transition at $\sim 19$ K which is generally robust to applied magnetic fields up to 9 T. Neutron diffraction experiments as a function of temperature and field allowed for the solution of the magnetic structure revising previous reports and suggesting a $\textbf{k}=(\frac{3}{2},0,0)$ ordering vector with the ostensibly QSL proximate ``zig-zag" magnetic order and a refined moment of 1.8(2)$\mu_B$ - consistent with the charge counting for a spin-1 Ni valence. Neutron powder diffraction up to 6 T showed no evidence of a metamagnetic transition but did reveal that the moment direction could be tuned within the $P_S\overline{1}$ magnetic space group indicating the possiblitiy for tuning to a \textbf{c} or \textbf{b}(\textbf{a}) polarized state in single crystals. 

Using inelastic neutron scattering the spin-Hamiltonian was studied. We found that both the extended Kitaev and isotropic-Heisenberg Hamiltonians were able to produce adequate simulations of the observed spectra. However, upon consideration the predictions of the zig-zag struture for a spin-1 Kitaev system with $J/K < 0$, the better capture of the spin-gap of the $\Gamma-K$ model, the inconsistent magnetic ground state of the ISO-model and its unusually large interlayer exchange all intimate a preference for the extended Kitaev Hamiltonian though additional inelastic measurements on co-aligned single crystal arrays are necessary to more firmly discriminate between these models. Using first principles calculations we showed that the zig-zag model is indeed the expected lowest energy magnetic order and that the inclusion of a U term was important to capture the experimental \Tn\ suggesting a highly localized system. Finally, looking to decrease the $J/K$ ratio we suggest alloying Pd on the Ni site which will both increase the spin-orbit coupling and thus strengthen the Kitaev interaction as well as reduce the $J$ due to enhancing the itineracy giving a road map to push \KNAO\ closer to the Kitaev state. Our work presents a rare example of a spin-1 honeycomb system which generally shows the possibility of proximity to a Kitaev state thus allowing for study of the associated physics in a higher-spin model than the canonical spin-$\frac{1}{2}$ quantum spin liquid.

\begin{acknowledgments}

The part of the research conducted at the High Flux Isotope Reactor and Spallation Neutron Source of Oak Ridge National Laboratory was sponsored by the Scientific User Facilities Division, Office of Basic Energy Sciences (BES), U.S. Department of Energy (DOE). The research is supported by the U.S. DOE, BES, Materials Science and Engineering Division. This research used resources at the Missouri University Research Reactor (MURR). This work was supported in part by a University of Missouri Research Council Grant (Grant Number: URC-22-021).

\end{acknowledgments}


\begin{thebibliography}{81}%
\makeatletter
\providecommand \@ifxundefined [1]{%
 \@ifx{#1\undefined}
}%
\providecommand \@ifnum [1]{%
 \ifnum #1\expandafter \@firstoftwo
 \else \expandafter \@secondoftwo
 \fi
}%
\providecommand \@ifx [1]{%
 \ifx #1\expandafter \@firstoftwo
 \else \expandafter \@secondoftwo
 \fi
}%
\providecommand \natexlab [1]{#1}%
\providecommand \enquote  [1]{``#1''}%
\providecommand \bibnamefont  [1]{#1}%
\providecommand \bibfnamefont [1]{#1}%
\providecommand \citenamefont [1]{#1}%
\providecommand \href@noop [0]{\@secondoftwo}%
\providecommand \href [0]{\begingroup \@sanitize@url \@href}%
\providecommand \@href[1]{\@@startlink{#1}\@@href}%
\providecommand \@@href[1]{\endgroup#1\@@endlink}%
\providecommand \@sanitize@url [0]{\catcode `\\12\catcode `\$12\catcode
  `\&12\catcode `\#12\catcode `\^12\catcode `\_12\catcode `\%12\relax}%
\providecommand \@@startlink[1]{}%
\providecommand \@@endlink[0]{}%
\providecommand \url  [0]{\begingroup\@sanitize@url \@url }%
\providecommand \@url [1]{\endgroup\@href {#1}{\urlprefix }}%
\providecommand \urlprefix  [0]{URL }%
\providecommand \Eprint [0]{\href }%
\providecommand \doibase [0]{https://doi.org/}%
\providecommand \selectlanguage [0]{\@gobble}%
\providecommand \bibinfo  [0]{\@secondoftwo}%
\providecommand \bibfield  [0]{\@secondoftwo}%
\providecommand \translation [1]{[#1]}%
\providecommand \BibitemOpen [0]{}%
\providecommand \bibitemStop [0]{}%
\providecommand \bibitemNoStop [0]{.\EOS\space}%
\providecommand \EOS [0]{\spacefactor3000\relax}%
\providecommand \BibitemShut  [1]{\csname bibitem#1\endcsname}%
\let\auto@bib@innerbib\@empty
\bibitem [{\citenamefont {Kitaev}(2006)}]{Kitaev2006}%
  \BibitemOpen
  \bibfield  {author} {\bibinfo {author} {\bibfnamefont {A.}~\bibnamefont
  {Kitaev}},\ }\bibfield  {title} {\bibinfo {title} {Anyons in an exactly
  solved model and beyond},\ }\href@noop {} {\bibfield  {journal} {\bibinfo
  {journal} {Ann. Phys.}\ }\textbf {\bibinfo {volume} {321}},\ \bibinfo {pages}
  {2} (\bibinfo {year} {2006})}\BibitemShut {NoStop}%
\bibitem [{\citenamefont {Savary}\ and\ \citenamefont
  {Balents}(2016)}]{Savary2016}%
  \BibitemOpen
  \bibfield  {author} {\bibinfo {author} {\bibfnamefont {L.}~\bibnamefont
  {Savary}}\ and\ \bibinfo {author} {\bibfnamefont {L.}~\bibnamefont
  {Balents}},\ }\bibfield  {title} {\bibinfo {title} {Quantum spin liquids: a
  review},\ }\href@noop {} {\bibfield  {journal} {\bibinfo  {journal} {Reports
  on Progress in Physics}\ }\textbf {\bibinfo {volume} {80}},\ \bibinfo {pages}
  {016502} (\bibinfo {year} {2016})}\BibitemShut {NoStop}%
\bibitem [{\citenamefont {Chamorro}\ \emph {et~al.}(2020)\citenamefont
  {Chamorro}, \citenamefont {McQueen},\ and\ \citenamefont
  {Tran}}]{Chamorro2020}%
  \BibitemOpen
  \bibfield  {author} {\bibinfo {author} {\bibfnamefont {J.~R.}\ \bibnamefont
  {Chamorro}}, \bibinfo {author} {\bibfnamefont {T.~M.}\ \bibnamefont
  {McQueen}},\ and\ \bibinfo {author} {\bibfnamefont {T.~T.}\ \bibnamefont
  {Tran}},\ }\bibfield  {title} {\bibinfo {title} {Chemistry of quantum spin
  liquids},\ }\href@noop {} {\bibfield  {journal} {\bibinfo  {journal} {Chem.
  Rev.}\ }\textbf {\bibinfo {volume} {121}},\ \bibinfo {pages} {2898} (\bibinfo
  {year} {2020})}\BibitemShut {NoStop}%
\bibitem [{\citenamefont {Trebst}\ and\ \citenamefont
  {Hickey}(2022)}]{Trebst2022}%
  \BibitemOpen
  \bibfield  {author} {\bibinfo {author} {\bibfnamefont {S.}~\bibnamefont
  {Trebst}}\ and\ \bibinfo {author} {\bibfnamefont {C.}~\bibnamefont
  {Hickey}},\ }\bibfield  {title} {\bibinfo {title} {Kitaev materials},\ }\href
  {https://doi.org/https://doi.org/10.1016/j.physrep.2021.11.003} {\bibfield
  {journal} {\bibinfo  {journal} {Phys. Rep.}\ }\textbf {\bibinfo {volume}
  {950}},\ \bibinfo {pages} {1} (\bibinfo {year} {2022})},\ \bibinfo {note}
  {kitaev materials}\BibitemShut {NoStop}%
\bibitem [{\citenamefont {Broholm}\ \emph {et~al.}(2020)\citenamefont
  {Broholm}, \citenamefont {Cava}, \citenamefont {Kivelson}, \citenamefont
  {Nocera}, \citenamefont {Norman},\ and\ \citenamefont
  {Senthil}}]{Broholm2020}%
  \BibitemOpen
  \bibfield  {author} {\bibinfo {author} {\bibfnamefont {C.}~\bibnamefont
  {Broholm}}, \bibinfo {author} {\bibfnamefont {R.}~\bibnamefont {Cava}},
  \bibinfo {author} {\bibfnamefont {S.}~\bibnamefont {Kivelson}}, \bibinfo
  {author} {\bibfnamefont {D.}~\bibnamefont {Nocera}}, \bibinfo {author}
  {\bibfnamefont {M.}~\bibnamefont {Norman}},\ and\ \bibinfo {author}
  {\bibfnamefont {T.}~\bibnamefont {Senthil}},\ }\bibfield  {title} {\bibinfo
  {title} {Quantum spin liquids},\ }\href@noop {} {\bibfield  {journal}
  {\bibinfo  {journal} {Science}\ }\textbf {\bibinfo {volume} {367}},\ \bibinfo
  {pages} {eaay0668} (\bibinfo {year} {2020})}\BibitemShut {NoStop}%
\bibitem [{\citenamefont {Semeghini}\ \emph {et~al.}(2021)\citenamefont
  {Semeghini}, \citenamefont {Levine}, \citenamefont {Keesling}, \citenamefont
  {Ebadi}, \citenamefont {Wang}, \citenamefont {Bluvstein}, \citenamefont
  {Verresen}, \citenamefont {Pichler}, \citenamefont {Kalinowski},
  \citenamefont {Samajdar} \emph {et~al.}}]{Semeghini2021}%
  \BibitemOpen
  \bibfield  {author} {\bibinfo {author} {\bibfnamefont {G.}~\bibnamefont
  {Semeghini}}, \bibinfo {author} {\bibfnamefont {H.}~\bibnamefont {Levine}},
  \bibinfo {author} {\bibfnamefont {A.}~\bibnamefont {Keesling}}, \bibinfo
  {author} {\bibfnamefont {S.}~\bibnamefont {Ebadi}}, \bibinfo {author}
  {\bibfnamefont {T.~T.}\ \bibnamefont {Wang}}, \bibinfo {author}
  {\bibfnamefont {D.}~\bibnamefont {Bluvstein}}, \bibinfo {author}
  {\bibfnamefont {R.}~\bibnamefont {Verresen}}, \bibinfo {author}
  {\bibfnamefont {H.}~\bibnamefont {Pichler}}, \bibinfo {author} {\bibfnamefont
  {M.}~\bibnamefont {Kalinowski}}, \bibinfo {author} {\bibfnamefont
  {R.}~\bibnamefont {Samajdar}}, \emph {et~al.},\ }\bibfield  {title} {\bibinfo
  {title} {Probing topological spin liquids on a programmable quantum
  simulator},\ }\href@noop {} {\bibfield  {journal} {\bibinfo  {journal}
  {Science}\ }\textbf {\bibinfo {volume} {374}},\ \bibinfo {pages} {1242}
  (\bibinfo {year} {2021})}\BibitemShut {NoStop}%
\bibitem [{\citenamefont {Takagi}\ \emph {et~al.}(2019)\citenamefont {Takagi},
  \citenamefont {Takayama}, \citenamefont {Jackeli}, \citenamefont
  {Khaliullin},\ and\ \citenamefont {Nagler}}]{Takagi2019}%
  \BibitemOpen
  \bibfield  {author} {\bibinfo {author} {\bibfnamefont {H.}~\bibnamefont
  {Takagi}}, \bibinfo {author} {\bibfnamefont {T.}~\bibnamefont {Takayama}},
  \bibinfo {author} {\bibfnamefont {G.}~\bibnamefont {Jackeli}}, \bibinfo
  {author} {\bibfnamefont {G.}~\bibnamefont {Khaliullin}},\ and\ \bibinfo
  {author} {\bibfnamefont {S.~E.}\ \bibnamefont {Nagler}},\ }\bibfield  {title}
  {\bibinfo {title} {{Concept and realization of Kitaev quantum spin
  liquids}},\ }\href@noop {} {\bibfield  {journal} {\bibinfo  {journal} {Nat.
  Rev. Phys.}\ }\textbf {\bibinfo {volume} {1}},\ \bibinfo {pages} {264}
  (\bibinfo {year} {2019})}\BibitemShut {NoStop}%
\bibitem [{\citenamefont {Banerjee}\ \emph {et~al.}(2016)\citenamefont
  {Banerjee}, \citenamefont {Bridges}, \citenamefont {Yan}, \citenamefont
  {Aczel}, \citenamefont {Li}, \citenamefont {Stone}, \citenamefont {Granroth},
  \citenamefont {Lumsden}, \citenamefont {Yiu}, \citenamefont {Knolle} \emph
  {et~al.}}]{Banerjee2016}%
  \BibitemOpen
  \bibfield  {author} {\bibinfo {author} {\bibfnamefont {A.}~\bibnamefont
  {Banerjee}}, \bibinfo {author} {\bibfnamefont {C.}~\bibnamefont {Bridges}},
  \bibinfo {author} {\bibfnamefont {J.-Q.}\ \bibnamefont {Yan}}, \bibinfo
  {author} {\bibfnamefont {A.}~\bibnamefont {Aczel}}, \bibinfo {author}
  {\bibfnamefont {L.}~\bibnamefont {Li}}, \bibinfo {author} {\bibfnamefont
  {M.}~\bibnamefont {Stone}}, \bibinfo {author} {\bibfnamefont
  {G.}~\bibnamefont {Granroth}}, \bibinfo {author} {\bibfnamefont
  {M.}~\bibnamefont {Lumsden}}, \bibinfo {author} {\bibfnamefont
  {Y.}~\bibnamefont {Yiu}}, \bibinfo {author} {\bibfnamefont {J.}~\bibnamefont
  {Knolle}}, \emph {et~al.},\ }\bibfield  {title} {\bibinfo {title} {Proximate
  kitaev quantum spin liquid behaviour in a honeycomb magnet},\ }\href@noop {}
  {\bibfield  {journal} {\bibinfo  {journal} {Nature materials}\ }\textbf
  {\bibinfo {volume} {15}},\ \bibinfo {pages} {733} (\bibinfo {year}
  {2016})}\BibitemShut {NoStop}%
\bibitem [{\citenamefont {Banerjee}\ \emph {et~al.}(2017)\citenamefont
  {Banerjee}, \citenamefont {Yan}, \citenamefont {Knolle}, \citenamefont
  {Bridges}, \citenamefont {Stone}, \citenamefont {Lumsden}, \citenamefont
  {Mandrus}, \citenamefont {Tennant}, \citenamefont {Moessner},\ and\
  \citenamefont {Nagler}}]{Banerjee2017}%
  \BibitemOpen
  \bibfield  {author} {\bibinfo {author} {\bibfnamefont {A.}~\bibnamefont
  {Banerjee}}, \bibinfo {author} {\bibfnamefont {J.}~\bibnamefont {Yan}},
  \bibinfo {author} {\bibfnamefont {J.}~\bibnamefont {Knolle}}, \bibinfo
  {author} {\bibfnamefont {C.~A.}\ \bibnamefont {Bridges}}, \bibinfo {author}
  {\bibfnamefont {M.~B.}\ \bibnamefont {Stone}}, \bibinfo {author}
  {\bibfnamefont {M.~D.}\ \bibnamefont {Lumsden}}, \bibinfo {author}
  {\bibfnamefont {D.~G.}\ \bibnamefont {Mandrus}}, \bibinfo {author}
  {\bibfnamefont {D.~A.}\ \bibnamefont {Tennant}}, \bibinfo {author}
  {\bibfnamefont {R.}~\bibnamefont {Moessner}},\ and\ \bibinfo {author}
  {\bibfnamefont {S.~E.}\ \bibnamefont {Nagler}},\ }\bibfield  {title}
  {\bibinfo {title} {Neutron scattering in the proximate quantum spin liquid
  $\alpha$-rucl3},\ }\href@noop {} {\bibfield  {journal} {\bibinfo  {journal}
  {Science}\ }\textbf {\bibinfo {volume} {356}},\ \bibinfo {pages} {1055}
  (\bibinfo {year} {2017})}\BibitemShut {NoStop}%
\bibitem [{\citenamefont {Yokoi}\ \emph {et~al.}(2021)\citenamefont {Yokoi},
  \citenamefont {Ma}, \citenamefont {Kasahara}, \citenamefont {Kasahara},
  \citenamefont {Shibauchi}, \citenamefont {Kurita}, \citenamefont {Tanaka},
  \citenamefont {Nasu}, \citenamefont {Motome}, \citenamefont {Hickey} \emph
  {et~al.}}]{Yokoi2021}%
  \BibitemOpen
  \bibfield  {author} {\bibinfo {author} {\bibfnamefont {T.}~\bibnamefont
  {Yokoi}}, \bibinfo {author} {\bibfnamefont {S.}~\bibnamefont {Ma}}, \bibinfo
  {author} {\bibfnamefont {Y.}~\bibnamefont {Kasahara}}, \bibinfo {author}
  {\bibfnamefont {S.}~\bibnamefont {Kasahara}}, \bibinfo {author}
  {\bibfnamefont {T.}~\bibnamefont {Shibauchi}}, \bibinfo {author}
  {\bibfnamefont {N.}~\bibnamefont {Kurita}}, \bibinfo {author} {\bibfnamefont
  {H.}~\bibnamefont {Tanaka}}, \bibinfo {author} {\bibfnamefont
  {J.}~\bibnamefont {Nasu}}, \bibinfo {author} {\bibfnamefont {Y.}~\bibnamefont
  {Motome}}, \bibinfo {author} {\bibfnamefont {C.}~\bibnamefont {Hickey}},
  \emph {et~al.},\ }\bibfield  {title} {\bibinfo {title} {{Half-integer
  quantized anomalous thermal Hall effect in the Kitaev material candidate
  $\alpha$-RuCl$_3$}},\ }\href@noop {} {\bibfield  {journal} {\bibinfo
  {journal} {Science}\ }\textbf {\bibinfo {volume} {373}},\ \bibinfo {pages}
  {568} (\bibinfo {year} {2021})}\BibitemShut {NoStop}%
\bibitem [{\citenamefont {Bruin}\ \emph {et~al.}(2022)\citenamefont {Bruin},
  \citenamefont {Claus}, \citenamefont {Matsumoto}, \citenamefont {Kurita},
  \citenamefont {Tanaka},\ and\ \citenamefont {Takagi}}]{Bruin2022}%
  \BibitemOpen
  \bibfield  {author} {\bibinfo {author} {\bibfnamefont {J.}~\bibnamefont
  {Bruin}}, \bibinfo {author} {\bibfnamefont {R.}~\bibnamefont {Claus}},
  \bibinfo {author} {\bibfnamefont {Y.}~\bibnamefont {Matsumoto}}, \bibinfo
  {author} {\bibfnamefont {N.}~\bibnamefont {Kurita}}, \bibinfo {author}
  {\bibfnamefont {H.}~\bibnamefont {Tanaka}},\ and\ \bibinfo {author}
  {\bibfnamefont {H.}~\bibnamefont {Takagi}},\ }\bibfield  {title} {\bibinfo
  {title} {{Robustness of the thermal Hall effect close to half-quantization in
  $\alpha$-RuCl$_3$}},\ }\href@noop {} {\bibfield  {journal} {\bibinfo
  {journal} {Nat. Phys.}\ }\textbf {\bibinfo {volume} {18}},\ \bibinfo {pages}
  {401} (\bibinfo {year} {2022})}\BibitemShut {NoStop}%
\bibitem [{\citenamefont {Anderson}(1987)}]{Anderson1987}%
  \BibitemOpen
  \bibfield  {author} {\bibinfo {author} {\bibfnamefont {P.}~\bibnamefont
  {Anderson}},\ }\bibfield  {title} {\bibinfo {title} {The resonating valence
  bond state in la2cuo4 and superconductivity},\ }\href
  {https://doi.org/10.1126/science.235.4793.1196} {\bibfield  {journal}
  {\bibinfo  {journal} {Science}\ }\textbf {\bibinfo {volume} {235}},\ \bibinfo
  {pages} {1196} (\bibinfo {year} {1987})}\BibitemShut {NoStop}%
\bibitem [{\citenamefont {Wen}(2002)}]{Wen2002}%
  \BibitemOpen
  \bibfield  {author} {\bibinfo {author} {\bibfnamefont {X.-G.}\ \bibnamefont
  {Wen}},\ }\bibfield  {title} {\bibinfo {title} {Quantum orders and symmetric
  spin liquids},\ }\href {https://doi.org/10.1103/PhysRevB.65.165113}
  {\bibfield  {journal} {\bibinfo  {journal} {Phys. Rev. B}\ }\textbf {\bibinfo
  {volume} {65}},\ \bibinfo {pages} {165113} (\bibinfo {year}
  {2002})}\BibitemShut {NoStop}%
\bibitem [{\citenamefont {Chamon}\ \emph {et~al.}(2017)\citenamefont {Chamon},
  \citenamefont {Goerbig}, \citenamefont {Moessner},\ and\ \citenamefont
  {Cugliandolo}}]{Chamon2017}%
  \BibitemOpen
  \bibfield  {author} {\bibinfo {author} {\bibfnamefont {C.}~\bibnamefont
  {Chamon}}, \bibinfo {author} {\bibfnamefont {M.~O.}\ \bibnamefont {Goerbig}},
  \bibinfo {author} {\bibfnamefont {R.}~\bibnamefont {Moessner}},\ and\
  \bibinfo {author} {\bibfnamefont {L.~F.}\ \bibnamefont {Cugliandolo}},\
  }\href@noop {} {\emph {\bibinfo {title} {Topological Aspects of Condensed
  Matter Physics: {\'E}cole de Physique Des Houches, Session CIII, 4-29 August
  2014}}},\ Vol.\ \bibinfo {volume} {103}\ (\bibinfo  {publisher} {Oxford
  University Press},\ \bibinfo {year} {2017})\BibitemShut {NoStop}%
\bibitem [{\citenamefont {Senthil}\ \emph {et~al.}(2003)\citenamefont
  {Senthil}, \citenamefont {Sachdev},\ and\ \citenamefont
  {Vojta}}]{Senthil2003}%
  \BibitemOpen
  \bibfield  {author} {\bibinfo {author} {\bibfnamefont {T.}~\bibnamefont
  {Senthil}}, \bibinfo {author} {\bibfnamefont {S.}~\bibnamefont {Sachdev}},\
  and\ \bibinfo {author} {\bibfnamefont {M.}~\bibnamefont {Vojta}},\ }\bibfield
   {title} {\bibinfo {title} {{Fractionalized Fermi Liquids}},\ }\href
  {https://doi.org/10.1103/PhysRevLett.90.216403} {\bibfield  {journal}
  {\bibinfo  {journal} {Phys. Rev. Lett.}\ }\textbf {\bibinfo {volume} {90}},\
  \bibinfo {pages} {216403} (\bibinfo {year} {2003})}\BibitemShut {NoStop}%
\bibitem [{\citenamefont {Kitaev}(2003)}]{Kitaev2003}%
  \BibitemOpen
  \bibfield  {author} {\bibinfo {author} {\bibfnamefont {A.~Y.}\ \bibnamefont
  {Kitaev}},\ }\bibfield  {title} {\bibinfo {title} {Fault-tolerant quantum
  computation by anyons},\ }\href@noop {} {\bibfield  {journal} {\bibinfo
  {journal} {Ann. Phys.}\ }\textbf {\bibinfo {volume} {303}},\ \bibinfo {pages}
  {2} (\bibinfo {year} {2003})}\BibitemShut {NoStop}%
\bibitem [{\citenamefont {Wen}\ \emph {et~al.}(2019)\citenamefont {Wen},
  \citenamefont {Yu}, \citenamefont {Li}, \citenamefont {Yu},\ and\
  \citenamefont {Li}}]{Wen2019}%
  \BibitemOpen
  \bibfield  {author} {\bibinfo {author} {\bibfnamefont {J.}~\bibnamefont
  {Wen}}, \bibinfo {author} {\bibfnamefont {S.-L.}\ \bibnamefont {Yu}},
  \bibinfo {author} {\bibfnamefont {S.}~\bibnamefont {Li}}, \bibinfo {author}
  {\bibfnamefont {W.}~\bibnamefont {Yu}},\ and\ \bibinfo {author}
  {\bibfnamefont {J.-X.}\ \bibnamefont {Li}},\ }\bibfield  {title} {\bibinfo
  {title} {Experimental identification of quantum spin liquids},\ }\href@noop
  {} {\bibfield  {journal} {\bibinfo  {journal} {npj Quantum Materials}\
  }\textbf {\bibinfo {volume} {4}},\ \bibinfo {pages} {1} (\bibinfo {year}
  {2019})}\BibitemShut {NoStop}%
\bibitem [{\citenamefont {Clark}\ and\ \citenamefont
  {Abdeldaim}(2021)}]{Clark2021}%
  \BibitemOpen
  \bibfield  {author} {\bibinfo {author} {\bibfnamefont {L.}~\bibnamefont
  {Clark}}\ and\ \bibinfo {author} {\bibfnamefont {A.~H.}\ \bibnamefont
  {Abdeldaim}},\ }\bibfield  {title} {\bibinfo {title} {Quantum spin liquids
  from a materials perspective},\ }\href@noop {} {\bibfield  {journal}
  {\bibinfo  {journal} {Annual Review of Materials Research}\ }\textbf
  {\bibinfo {volume} {51}},\ \bibinfo {pages} {495} (\bibinfo {year}
  {2021})}\BibitemShut {NoStop}%
\bibitem [{\citenamefont {Knolle}\ and\ \citenamefont
  {Moessner}(2019)}]{Knolle2019}%
  \BibitemOpen
  \bibfield  {author} {\bibinfo {author} {\bibfnamefont {J.}~\bibnamefont
  {Knolle}}\ and\ \bibinfo {author} {\bibfnamefont {R.}~\bibnamefont
  {Moessner}},\ }\bibfield  {title} {\bibinfo {title} {A field guide to spin
  liquids},\ }\href@noop {} {\bibfield  {journal} {\bibinfo  {journal} {Annu.
  Rev. Condens. Matter Phys.}\ }\textbf {\bibinfo {volume} {10}},\ \bibinfo
  {pages} {451} (\bibinfo {year} {2019})}\BibitemShut {NoStop}%
\bibitem [{\citenamefont {Jackeli}\ and\ \citenamefont
  {Khaliullin}(2009)}]{Jackeli2009}%
  \BibitemOpen
  \bibfield  {author} {\bibinfo {author} {\bibfnamefont {G.}~\bibnamefont
  {Jackeli}}\ and\ \bibinfo {author} {\bibfnamefont {G.}~\bibnamefont
  {Khaliullin}},\ }\bibfield  {title} {\bibinfo {title} {Mott insulators in the
  strong spin-orbit coupling limit: From heisenberg to a quantum compass and
  kitaev models},\ }\href {https://doi.org/10.1103/PhysRevLett.102.017205}
  {\bibfield  {journal} {\bibinfo  {journal} {Phys. Rev. Lett.}\ }\textbf
  {\bibinfo {volume} {102}},\ \bibinfo {pages} {017205} (\bibinfo {year}
  {2009})}\BibitemShut {NoStop}%
\bibitem [{\citenamefont {Rao}\ \emph {et~al.}(2021)\citenamefont {Rao},
  \citenamefont {Hussain}, \citenamefont {Huang}, \citenamefont {Chu},
  \citenamefont {Li}, \citenamefont {Zhao}, \citenamefont {Dun}, \citenamefont
  {Choi}, \citenamefont {Asaba}, \citenamefont {Chen} \emph
  {et~al.}}]{Rao2021}%
  \BibitemOpen
  \bibfield  {author} {\bibinfo {author} {\bibfnamefont {X.}~\bibnamefont
  {Rao}}, \bibinfo {author} {\bibfnamefont {G.}~\bibnamefont {Hussain}},
  \bibinfo {author} {\bibfnamefont {Q.}~\bibnamefont {Huang}}, \bibinfo
  {author} {\bibfnamefont {W.}~\bibnamefont {Chu}}, \bibinfo {author}
  {\bibfnamefont {N.}~\bibnamefont {Li}}, \bibinfo {author} {\bibfnamefont
  {X.}~\bibnamefont {Zhao}}, \bibinfo {author} {\bibfnamefont {Z.}~\bibnamefont
  {Dun}}, \bibinfo {author} {\bibfnamefont {E.}~\bibnamefont {Choi}}, \bibinfo
  {author} {\bibfnamefont {T.}~\bibnamefont {Asaba}}, \bibinfo {author}
  {\bibfnamefont {L.}~\bibnamefont {Chen}}, \emph {et~al.},\ }\bibfield
  {title} {\bibinfo {title} {{Survival of itinerant excitations and quantum
  spin state transitions in YbMgGaO$_4$ with chemical disorder}},\ }\href@noop
  {} {\bibfield  {journal} {\bibinfo  {journal} {Nat. Comm.}\ }\textbf
  {\bibinfo {volume} {12}},\ \bibinfo {pages} {1} (\bibinfo {year}
  {2021})}\BibitemShut {NoStop}%
\bibitem [{\citenamefont {Huang}\ \emph {et~al.}(2021)\citenamefont {Huang},
  \citenamefont {Xu}, \citenamefont {Wang}, \citenamefont {Zhao}, \citenamefont
  {Tu}, \citenamefont {Ni}, \citenamefont {Wang}, \citenamefont {Pan},
  \citenamefont {Fu}, \citenamefont {Hao}, \citenamefont {Liu}, \citenamefont
  {Mei},\ and\ \citenamefont {Li}}]{Huang2021}%
  \BibitemOpen
  \bibfield  {author} {\bibinfo {author} {\bibfnamefont {Y.~Y.}\ \bibnamefont
  {Huang}}, \bibinfo {author} {\bibfnamefont {Y.}~\bibnamefont {Xu}}, \bibinfo
  {author} {\bibfnamefont {L.}~\bibnamefont {Wang}}, \bibinfo {author}
  {\bibfnamefont {C.~C.}\ \bibnamefont {Zhao}}, \bibinfo {author}
  {\bibfnamefont {C.~P.}\ \bibnamefont {Tu}}, \bibinfo {author} {\bibfnamefont
  {J.~M.}\ \bibnamefont {Ni}}, \bibinfo {author} {\bibfnamefont {L.~S.}\
  \bibnamefont {Wang}}, \bibinfo {author} {\bibfnamefont {B.~L.}\ \bibnamefont
  {Pan}}, \bibinfo {author} {\bibfnamefont {Y.}~\bibnamefont {Fu}}, \bibinfo
  {author} {\bibfnamefont {Z.}~\bibnamefont {Hao}}, \bibinfo {author}
  {\bibfnamefont {C.}~\bibnamefont {Liu}}, \bibinfo {author} {\bibfnamefont
  {J.-W.}\ \bibnamefont {Mei}},\ and\ \bibinfo {author} {\bibfnamefont {S.~Y.}\
  \bibnamefont {Li}},\ }\bibfield  {title} {\bibinfo {title} {{Heat Transport
  in Herbertsmithite: Can a Quantum Spin Liquid Survive Disorder?}},\ }\href
  {https://doi.org/10.1103/PhysRevLett.127.267202} {\bibfield  {journal}
  {\bibinfo  {journal} {Phys. Rev. Lett.}\ }\textbf {\bibinfo {volume} {127}},\
  \bibinfo {pages} {267202} (\bibinfo {year} {2021})}\BibitemShut {NoStop}%
\bibitem [{\citenamefont {Zhu}\ \emph {et~al.}(2017)\citenamefont {Zhu},
  \citenamefont {Maksimov}, \citenamefont {White},\ and\ \citenamefont
  {Chernyshev}}]{Zhu2017}%
  \BibitemOpen
  \bibfield  {author} {\bibinfo {author} {\bibfnamefont {Z.}~\bibnamefont
  {Zhu}}, \bibinfo {author} {\bibfnamefont {P.~A.}\ \bibnamefont {Maksimov}},
  \bibinfo {author} {\bibfnamefont {S.~R.}\ \bibnamefont {White}},\ and\
  \bibinfo {author} {\bibfnamefont {A.~L.}\ \bibnamefont {Chernyshev}},\
  }\bibfield  {title} {\bibinfo {title} {{Disorder-Induced Mimicry of a Spin
  Liquid in ${\mathrm{YbMgGaO}}_{4}$}},\ }\href
  {https://doi.org/10.1103/PhysRevLett.119.157201} {\bibfield  {journal}
  {\bibinfo  {journal} {Phys. Rev. Lett.}\ }\textbf {\bibinfo {volume} {119}},\
  \bibinfo {pages} {157201} (\bibinfo {year} {2017})}\BibitemShut {NoStop}%
\bibitem [{\citenamefont {Ma}\ \emph {et~al.}(2020)\citenamefont {Ma},
  \citenamefont {Dong}, \citenamefont {Wu}, \citenamefont {Zhu}, \citenamefont
  {Bao}, \citenamefont {Cai}, \citenamefont {Wang}, \citenamefont {Shangguan},
  \citenamefont {Wang}, \citenamefont {Ran}, \citenamefont {Yu}, \citenamefont
  {Deng}, \citenamefont {Mole}, \citenamefont {Li}, \citenamefont {Yu},
  \citenamefont {Li},\ and\ \citenamefont {Wen}}]{Ma2020q}%
  \BibitemOpen
  \bibfield  {author} {\bibinfo {author} {\bibfnamefont {Z.}~\bibnamefont
  {Ma}}, \bibinfo {author} {\bibfnamefont {Z.-Y.}\ \bibnamefont {Dong}},
  \bibinfo {author} {\bibfnamefont {S.}~\bibnamefont {Wu}}, \bibinfo {author}
  {\bibfnamefont {Y.}~\bibnamefont {Zhu}}, \bibinfo {author} {\bibfnamefont
  {S.}~\bibnamefont {Bao}}, \bibinfo {author} {\bibfnamefont {Z.}~\bibnamefont
  {Cai}}, \bibinfo {author} {\bibfnamefont {W.}~\bibnamefont {Wang}}, \bibinfo
  {author} {\bibfnamefont {Y.}~\bibnamefont {Shangguan}}, \bibinfo {author}
  {\bibfnamefont {J.}~\bibnamefont {Wang}}, \bibinfo {author} {\bibfnamefont
  {K.}~\bibnamefont {Ran}}, \bibinfo {author} {\bibfnamefont {D.}~\bibnamefont
  {Yu}}, \bibinfo {author} {\bibfnamefont {G.}~\bibnamefont {Deng}}, \bibinfo
  {author} {\bibfnamefont {R.~A.}\ \bibnamefont {Mole}}, \bibinfo {author}
  {\bibfnamefont {H.-F.}\ \bibnamefont {Li}}, \bibinfo {author} {\bibfnamefont
  {S.-L.}\ \bibnamefont {Yu}}, \bibinfo {author} {\bibfnamefont {J.-X.}\
  \bibnamefont {Li}},\ and\ \bibinfo {author} {\bibfnamefont {J.}~\bibnamefont
  {Wen}},\ }\bibfield  {title} {\bibinfo {title} {Disorder-induced
  spin-liquid-like behavior in kagome-lattice compounds},\ }\href
  {https://doi.org/10.1103/PhysRevB.102.224415} {\bibfield  {journal} {\bibinfo
   {journal} {Phys. Rev. B}\ }\textbf {\bibinfo {volume} {102}},\ \bibinfo
  {pages} {224415} (\bibinfo {year} {2020})}\BibitemShut {NoStop}%
\bibitem [{\citenamefont {Baskaran}\ \emph {et~al.}(2008)\citenamefont
  {Baskaran}, \citenamefont {Sen},\ and\ \citenamefont
  {Shankar}}]{Baskaran2008}%
  \BibitemOpen
  \bibfield  {author} {\bibinfo {author} {\bibfnamefont {G.}~\bibnamefont
  {Baskaran}}, \bibinfo {author} {\bibfnamefont {D.}~\bibnamefont {Sen}},\ and\
  \bibinfo {author} {\bibfnamefont {R.}~\bibnamefont {Shankar}},\ }\bibfield
  {title} {\bibinfo {title} {{Spin-$S$ Kitaev model: Classical ground states,
  order from disorder, and exact correlation functions}},\ }\href
  {https://doi.org/10.1103/PhysRevB.78.115116} {\bibfield  {journal} {\bibinfo
  {journal} {Phys. Rev. B}\ }\textbf {\bibinfo {volume} {78}},\ \bibinfo
  {pages} {115116} (\bibinfo {year} {2008})}\BibitemShut {NoStop}%
\bibitem [{\citenamefont {Stavropoulos}\ \emph {et~al.}(2019)\citenamefont
  {Stavropoulos}, \citenamefont {Pereira},\ and\ \citenamefont
  {Kee}}]{Stavropoulos2019}%
  \BibitemOpen
  \bibfield  {author} {\bibinfo {author} {\bibfnamefont {P.~P.}\ \bibnamefont
  {Stavropoulos}}, \bibinfo {author} {\bibfnamefont {D.}~\bibnamefont
  {Pereira}},\ and\ \bibinfo {author} {\bibfnamefont {H.-Y.}\ \bibnamefont
  {Kee}},\ }\bibfield  {title} {\bibinfo {title} {{Microscopic Mechanism for a
  Higher-Spin Kitaev Model}},\ }\href
  {https://doi.org/10.1103/PhysRevLett.123.037203} {\bibfield  {journal}
  {\bibinfo  {journal} {Phys. Rev. Lett.}\ }\textbf {\bibinfo {volume} {123}},\
  \bibinfo {pages} {037203} (\bibinfo {year} {2019})}\BibitemShut {NoStop}%
\bibitem [{\citenamefont {Khait}\ \emph {et~al.}(2021)\citenamefont {Khait},
  \citenamefont {Stavropoulos}, \citenamefont {Kee},\ and\ \citenamefont
  {Kim}}]{Khait2021}%
  \BibitemOpen
  \bibfield  {author} {\bibinfo {author} {\bibfnamefont {I.}~\bibnamefont
  {Khait}}, \bibinfo {author} {\bibfnamefont {P.~P.}\ \bibnamefont
  {Stavropoulos}}, \bibinfo {author} {\bibfnamefont {H.-Y.}\ \bibnamefont
  {Kee}},\ and\ \bibinfo {author} {\bibfnamefont {Y.~B.}\ \bibnamefont {Kim}},\
  }\bibfield  {title} {\bibinfo {title} {Characterizing spin-one kitaev quantum
  spin liquids},\ }\href {https://doi.org/10.1103/PhysRevResearch.3.013160}
  {\bibfield  {journal} {\bibinfo  {journal} {Phys. Rev. Research}\ }\textbf
  {\bibinfo {volume} {3}},\ \bibinfo {pages} {013160} (\bibinfo {year}
  {2021})}\BibitemShut {NoStop}%
\bibitem [{\citenamefont {Dong}\ and\ \citenamefont {Sheng}(2020)}]{Dong2020}%
  \BibitemOpen
  \bibfield  {author} {\bibinfo {author} {\bibfnamefont {X.-Y.}\ \bibnamefont
  {Dong}}\ and\ \bibinfo {author} {\bibfnamefont {D.~N.}\ \bibnamefont
  {Sheng}},\ }\bibfield  {title} {\bibinfo {title} {{Spin-1 Kitaev-Heisenberg
  model on a honeycomb lattice}},\ }\href
  {https://doi.org/10.1103/PhysRevB.102.121102} {\bibfield  {journal} {\bibinfo
   {journal} {Phys. Rev. B}\ }\textbf {\bibinfo {volume} {102}},\ \bibinfo
  {pages} {121102} (\bibinfo {year} {2020})}\BibitemShut {NoStop}%
\bibitem [{\citenamefont {Lee}\ \emph {et~al.}(2020)\citenamefont {Lee},
  \citenamefont {Kawashima},\ and\ \citenamefont {Kim}}]{Lee2020}%
  \BibitemOpen
  \bibfield  {author} {\bibinfo {author} {\bibfnamefont {H.-Y.}\ \bibnamefont
  {Lee}}, \bibinfo {author} {\bibfnamefont {N.}~\bibnamefont {Kawashima}},\
  and\ \bibinfo {author} {\bibfnamefont {Y.~B.}\ \bibnamefont {Kim}},\
  }\bibfield  {title} {\bibinfo {title} {{Tensor network wave function of $S=1$
  Kitaev spin liquids}},\ }\href
  {https://doi.org/10.1103/PhysRevResearch.2.033318} {\bibfield  {journal}
  {\bibinfo  {journal} {Phys. Rev. Research}\ }\textbf {\bibinfo {volume}
  {2}},\ \bibinfo {pages} {033318} (\bibinfo {year} {2020})}\BibitemShut
  {NoStop}%
\bibitem [{\citenamefont {Koga}\ \emph {et~al.}(2018)\citenamefont {Koga},
  \citenamefont {Tomishige},\ and\ \citenamefont {Nasu}}]{Koga2018}%
  \BibitemOpen
  \bibfield  {author} {\bibinfo {author} {\bibfnamefont {A.}~\bibnamefont
  {Koga}}, \bibinfo {author} {\bibfnamefont {H.}~\bibnamefont {Tomishige}},\
  and\ \bibinfo {author} {\bibfnamefont {J.}~\bibnamefont {Nasu}},\ }\bibfield
  {title} {\bibinfo {title} {{Ground-state and thermodynamic properties of an S
  = 1 Kitaev model}},\ }\href@noop {} {\bibfield  {journal} {\bibinfo
  {journal} {J. Phys. Soc. Jpn.}\ }\textbf {\bibinfo {volume} {87}},\ \bibinfo
  {pages} {063703} (\bibinfo {year} {2018})}\BibitemShut {NoStop}%
\bibitem [{\citenamefont {Koga}\ \emph {et~al.}(2020)\citenamefont {Koga},
  \citenamefont {Minakawa}, \citenamefont {Murakami},\ and\ \citenamefont
  {Nasu}}]{Koga2020}%
  \BibitemOpen
  \bibfield  {author} {\bibinfo {author} {\bibfnamefont {A.}~\bibnamefont
  {Koga}}, \bibinfo {author} {\bibfnamefont {T.}~\bibnamefont {Minakawa}},
  \bibinfo {author} {\bibfnamefont {Y.}~\bibnamefont {Murakami}},\ and\
  \bibinfo {author} {\bibfnamefont {J.}~\bibnamefont {Nasu}},\ }\bibfield
  {title} {\bibinfo {title} {{Spin transport in the quantum spin liquid state
  in the S= 1 Kitaev model: role of the fractionalized quasiparticles}},\
  }\href@noop {} {\bibfield  {journal} {\bibinfo  {journal} {J. Phys. Soc.
  Jpn.}\ }\textbf {\bibinfo {volume} {89}},\ \bibinfo {pages} {033701}
  (\bibinfo {year} {2020})}\BibitemShut {NoStop}%
\bibitem [{\citenamefont {Chen}\ \emph {et~al.}(2022)\citenamefont {Chen},
  \citenamefont {Genzor}, \citenamefont {Kim},\ and\ \citenamefont
  {Kao}}]{Chen2022}%
  \BibitemOpen
  \bibfield  {author} {\bibinfo {author} {\bibfnamefont {Y.-H.}\ \bibnamefont
  {Chen}}, \bibinfo {author} {\bibfnamefont {J.}~\bibnamefont {Genzor}},
  \bibinfo {author} {\bibfnamefont {Y.~B.}\ \bibnamefont {Kim}},\ and\ \bibinfo
  {author} {\bibfnamefont {Y.-J.}\ \bibnamefont {Kao}},\ }\bibfield  {title}
  {\bibinfo {title} {{Excitation spectrum of spin-1 Kitaev spin liquids}},\
  }\href {https://doi.org/10.1103/PhysRevB.105.L060403} {\bibfield  {journal}
  {\bibinfo  {journal} {Phys. Rev. B}\ }\textbf {\bibinfo {volume} {105}},\
  \bibinfo {pages} {L060403} (\bibinfo {year} {2022})}\BibitemShut {NoStop}%
\bibitem [{\citenamefont {Zhu}\ \emph {et~al.}(2020)\citenamefont {Zhu},
  \citenamefont {Weng},\ and\ \citenamefont {Sheng}}]{Zhu2020}%
  \BibitemOpen
  \bibfield  {author} {\bibinfo {author} {\bibfnamefont {Z.}~\bibnamefont
  {Zhu}}, \bibinfo {author} {\bibfnamefont {Z.-Y.}\ \bibnamefont {Weng}},\ and\
  \bibinfo {author} {\bibfnamefont {D.~N.}\ \bibnamefont {Sheng}},\ }\bibfield
  {title} {\bibinfo {title} {{Magnetic field induced spin liquids in $S=1$
  Kitaev honeycomb model}},\ }\href
  {https://doi.org/10.1103/PhysRevResearch.2.022047} {\bibfield  {journal}
  {\bibinfo  {journal} {Phys. Rev. Research}\ }\textbf {\bibinfo {volume}
  {2}},\ \bibinfo {pages} {022047} (\bibinfo {year} {2020})}\BibitemShut
  {NoStop}%
\bibitem [{\citenamefont {Badrtdinov}\ \emph {et~al.}(2021)\citenamefont
  {Badrtdinov}, \citenamefont {Ding}, \citenamefont {Ritter}, \citenamefont
  {Hembacher}, \citenamefont {Ahmed}, \citenamefont {Skourski},\ and\
  \citenamefont {Tsirlin}}]{Badrtdinov2021}%
  \BibitemOpen
  \bibfield  {author} {\bibinfo {author} {\bibfnamefont {D.~I.}\ \bibnamefont
  {Badrtdinov}}, \bibinfo {author} {\bibfnamefont {L.}~\bibnamefont {Ding}},
  \bibinfo {author} {\bibfnamefont {C.}~\bibnamefont {Ritter}}, \bibinfo
  {author} {\bibfnamefont {J.}~\bibnamefont {Hembacher}}, \bibinfo {author}
  {\bibfnamefont {N.}~\bibnamefont {Ahmed}}, \bibinfo {author} {\bibfnamefont
  {Y.}~\bibnamefont {Skourski}},\ and\ \bibinfo {author} {\bibfnamefont
  {A.~A.}\ \bibnamefont {Tsirlin}},\ }\bibfield  {title} {\bibinfo {title}
  {{$\mathrm{Mo}{\mathrm{P}}_{3}\mathrm{Si}{\mathrm{O}}_{11}$: A $4{d}^{3}$
  honeycomb antiferromagnet with disconnected octahedra}},\ }\href
  {https://doi.org/10.1103/PhysRevB.104.094428} {\bibfield  {journal} {\bibinfo
   {journal} {Phys. Rev. B}\ }\textbf {\bibinfo {volume} {104}},\ \bibinfo
  {pages} {094428} (\bibinfo {year} {2021})}\BibitemShut {NoStop}%
\bibitem [{\citenamefont {Zhou}\ \emph {et~al.}(2021)\citenamefont {Zhou},
  \citenamefont {Chen}, \citenamefont {Luo}, \citenamefont {Luo},\ and\
  \citenamefont {Zhao}}]{Zhou2021}%
  \BibitemOpen
  \bibfield  {author} {\bibinfo {author} {\bibfnamefont {Z.}~\bibnamefont
  {Zhou}}, \bibinfo {author} {\bibfnamefont {K.}~\bibnamefont {Chen}}, \bibinfo
  {author} {\bibfnamefont {Q.}~\bibnamefont {Luo}}, \bibinfo {author}
  {\bibfnamefont {H.-G.}\ \bibnamefont {Luo}},\ and\ \bibinfo {author}
  {\bibfnamefont {J.}~\bibnamefont {Zhao}},\ }\bibfield  {title} {\bibinfo
  {title} {{Strain-induced phase diagram of the $S=\frac{3}{2}$ Kitaev material
  ${\mathrm{CrSiTe}}_{3}$}},\ }\href
  {https://doi.org/10.1103/PhysRevB.104.214425} {\bibfield  {journal} {\bibinfo
   {journal} {Phys. Rev. B}\ }\textbf {\bibinfo {volume} {104}},\ \bibinfo
  {pages} {214425} (\bibinfo {year} {2021})}\BibitemShut {NoStop}%
\bibitem [{\citenamefont {Chen}\ \emph {et~al.}(2021)\citenamefont {Chen},
  \citenamefont {Chung}, \citenamefont {Stone}, \citenamefont {Kolesnikov},
  \citenamefont {Winn}, \citenamefont {Garlea}, \citenamefont {Abernathy},
  \citenamefont {Gao}, \citenamefont {Augustin}, \citenamefont {Santos},\ and\
  \citenamefont {Dai}}]{Chen2021}%
  \BibitemOpen
  \bibfield  {author} {\bibinfo {author} {\bibfnamefont {L.}~\bibnamefont
  {Chen}}, \bibinfo {author} {\bibfnamefont {J.-H.}\ \bibnamefont {Chung}},
  \bibinfo {author} {\bibfnamefont {M.~B.}\ \bibnamefont {Stone}}, \bibinfo
  {author} {\bibfnamefont {A.~I.}\ \bibnamefont {Kolesnikov}}, \bibinfo
  {author} {\bibfnamefont {B.}~\bibnamefont {Winn}}, \bibinfo {author}
  {\bibfnamefont {V.~O.}\ \bibnamefont {Garlea}}, \bibinfo {author}
  {\bibfnamefont {D.~L.}\ \bibnamefont {Abernathy}}, \bibinfo {author}
  {\bibfnamefont {B.}~\bibnamefont {Gao}}, \bibinfo {author} {\bibfnamefont
  {M.}~\bibnamefont {Augustin}}, \bibinfo {author} {\bibfnamefont {E.~J.~G.}\
  \bibnamefont {Santos}},\ and\ \bibinfo {author} {\bibfnamefont
  {P.}~\bibnamefont {Dai}},\ }\bibfield  {title} {\bibinfo {title} {{Magnetic
  Field Effect on Topological Spin Excitations in ${\mathrm{CrI}}_{3}$}},\
  }\href {https://doi.org/10.1103/PhysRevX.11.031047} {\bibfield  {journal}
  {\bibinfo  {journal} {Phys. Rev. X}\ }\textbf {\bibinfo {volume} {11}},\
  \bibinfo {pages} {031047} (\bibinfo {year} {2021})}\BibitemShut {NoStop}%
\bibitem [{\citenamefont {Samarakoon}\ \emph {et~al.}(2021)\citenamefont
  {Samarakoon}, \citenamefont {Chen}, \citenamefont {Zhou},\ and\ \citenamefont
  {Garlea}}]{Samarakoon2021}%
  \BibitemOpen
  \bibfield  {author} {\bibinfo {author} {\bibfnamefont {A.~M.}\ \bibnamefont
  {Samarakoon}}, \bibinfo {author} {\bibfnamefont {Q.}~\bibnamefont {Chen}},
  \bibinfo {author} {\bibfnamefont {H.}~\bibnamefont {Zhou}},\ and\ \bibinfo
  {author} {\bibfnamefont {V.~O.}\ \bibnamefont {Garlea}},\ }\bibfield  {title}
  {\bibinfo {title} {{Static and dynamic magnetic properties of honeycomb
  lattice antiferromagnets ${\mathrm{Na}}_{2}{M}_{2}{\mathrm{TeO}}_{6}$,
  $M=\mathrm{Co}$ and Ni}},\ }\href
  {https://doi.org/10.1103/PhysRevB.104.184415} {\bibfield  {journal} {\bibinfo
   {journal} {Phys. Rev. B}\ }\textbf {\bibinfo {volume} {104}},\ \bibinfo
  {pages} {184415} (\bibinfo {year} {2021})}\BibitemShut {NoStop}%
\bibitem [{\citenamefont {Bera}\ \emph {et~al.}(2022)\citenamefont {Bera},
  \citenamefont {Yusuf}, \citenamefont {Keller}, \citenamefont {Yokaichiya},\
  and\ \citenamefont {Stewart}}]{Bera2022}%
  \BibitemOpen
  \bibfield  {author} {\bibinfo {author} {\bibfnamefont {A.~K.}\ \bibnamefont
  {Bera}}, \bibinfo {author} {\bibfnamefont {S.~M.}\ \bibnamefont {Yusuf}},
  \bibinfo {author} {\bibfnamefont {L.}~\bibnamefont {Keller}}, \bibinfo
  {author} {\bibfnamefont {F.}~\bibnamefont {Yokaichiya}},\ and\ \bibinfo
  {author} {\bibfnamefont {J.~R.}\ \bibnamefont {Stewart}},\ }\bibfield
  {title} {\bibinfo {title} {{Magnetism of two-dimensional honeycomb layered
  ${\mathrm{Na}}_{2}{\mathrm{Ni}}_{2}\mathrm{Te}{\mathrm{O}}_{6}$ driven by
  intermediate Na-layer crystal structure}},\ }\href
  {https://doi.org/10.1103/PhysRevB.105.014410} {\bibfield  {journal} {\bibinfo
   {journal} {Phys. Rev. B}\ }\textbf {\bibinfo {volume} {105}},\ \bibinfo
  {pages} {014410} (\bibinfo {year} {2022})}\BibitemShut {NoStop}%
\bibitem [{\citenamefont {Lefran\ifmmode~\mbox{\c{c}}\else \c{c}\fi{}ois}\
  \emph {et~al.}(2016)\citenamefont {Lefran\ifmmode~\mbox{\c{c}}\else
  \c{c}\fi{}ois}, \citenamefont {Songvilay}, \citenamefont {Robert},
  \citenamefont {Nataf}, \citenamefont {Jordan}, \citenamefont {Chaix},
  \citenamefont {Colin}, \citenamefont {Lejay}, \citenamefont {Hadj-Azzem},
  \citenamefont {Ballou},\ and\ \citenamefont {Simonet}}]{Lefrancois2016}%
  \BibitemOpen
  \bibfield  {author} {\bibinfo {author} {\bibfnamefont {E.}~\bibnamefont
  {Lefran\ifmmode~\mbox{\c{c}}\else \c{c}\fi{}ois}}, \bibinfo {author}
  {\bibfnamefont {M.}~\bibnamefont {Songvilay}}, \bibinfo {author}
  {\bibfnamefont {J.}~\bibnamefont {Robert}}, \bibinfo {author} {\bibfnamefont
  {G.}~\bibnamefont {Nataf}}, \bibinfo {author} {\bibfnamefont
  {E.}~\bibnamefont {Jordan}}, \bibinfo {author} {\bibfnamefont
  {L.}~\bibnamefont {Chaix}}, \bibinfo {author} {\bibfnamefont {C.~V.}\
  \bibnamefont {Colin}}, \bibinfo {author} {\bibfnamefont {P.}~\bibnamefont
  {Lejay}}, \bibinfo {author} {\bibfnamefont {A.}~\bibnamefont {Hadj-Azzem}},
  \bibinfo {author} {\bibfnamefont {R.}~\bibnamefont {Ballou}},\ and\ \bibinfo
  {author} {\bibfnamefont {V.}~\bibnamefont {Simonet}},\ }\bibfield  {title}
  {\bibinfo {title} {{Magnetic properties of the honeycomb oxide
  ${\mathbf{Na}}_{2}{\mathbf{Co}}_{2}{\mathbf{TeO}}_{6}$}},\ }\href
  {https://doi.org/10.1103/PhysRevB.94.214416} {\bibfield  {journal} {\bibinfo
  {journal} {Phys. Rev. B}\ }\textbf {\bibinfo {volume} {94}},\ \bibinfo
  {pages} {214416} (\bibinfo {year} {2016})}\BibitemShut {NoStop}%
\bibitem [{\citenamefont {Kurbakov}\ \emph {et~al.}(2020)\citenamefont
  {Kurbakov}, \citenamefont {Korshunov}, \citenamefont {Podchezertsev},
  \citenamefont {Stratan}, \citenamefont {Raganyan},\ and\ \citenamefont
  {Zvereva}}]{Kurbakov2020}%
  \BibitemOpen
  \bibfield  {author} {\bibinfo {author} {\bibfnamefont {A.~I.}\ \bibnamefont
  {Kurbakov}}, \bibinfo {author} {\bibfnamefont {A.~N.}\ \bibnamefont
  {Korshunov}}, \bibinfo {author} {\bibfnamefont {S.~Y.}\ \bibnamefont
  {Podchezertsev}}, \bibinfo {author} {\bibfnamefont {M.~I.}\ \bibnamefont
  {Stratan}}, \bibinfo {author} {\bibfnamefont {G.~V.}\ \bibnamefont
  {Raganyan}},\ and\ \bibinfo {author} {\bibfnamefont {E.~A.}\ \bibnamefont
  {Zvereva}},\ }\bibfield  {title} {\bibinfo {title} {{Long-range and
  short-range ordering in 2D honeycomb-lattice magnet Na$_2$Ni$_2$TeO$_6$}},\
  }\href@noop {} {\bibfield  {journal} {\bibinfo  {journal} {J. Alloys Compd.}\
  }\textbf {\bibinfo {volume} {820}},\ \bibinfo {pages} {153354} (\bibinfo
  {year} {2020})}\BibitemShut {NoStop}%
\bibitem [{\citenamefont {Ladwig}\ and\ \citenamefont
  {Ziemer}(1979)}]{Ladwig1979}%
  \BibitemOpen
  \bibfield  {author} {\bibinfo {author} {\bibfnamefont {G.}~\bibnamefont
  {Ladwig}}\ and\ \bibinfo {author} {\bibfnamefont {B.}~\bibnamefont
  {Ziemer}},\ }\bibfield  {title} {\bibinfo {title} {{{\"U}ber das
  glimmerartige Kaliumnickel (II)-monoarsenat KNiAsO$_4$}},\ }\href@noop {}
  {\bibfield  {journal} {\bibinfo  {journal} {Z.anorg. allg. Chem.}\ }\textbf
  {\bibinfo {volume} {457}},\ \bibinfo {pages} {143} (\bibinfo {year}
  {1979})}\BibitemShut {NoStop}%
\bibitem [{\citenamefont {Buckley}\ \emph {et~al.}(1987)\citenamefont
  {Buckley}, \citenamefont {Bramwell}, \citenamefont {Visser},\ and\
  \citenamefont {Day}}]{Buckley1987}%
  \BibitemOpen
  \bibfield  {author} {\bibinfo {author} {\bibfnamefont {A.}~\bibnamefont
  {Buckley}}, \bibinfo {author} {\bibfnamefont {S.}~\bibnamefont {Bramwell}},
  \bibinfo {author} {\bibfnamefont {D.}~\bibnamefont {Visser}},\ and\ \bibinfo
  {author} {\bibfnamefont {P.}~\bibnamefont {Day}},\ }\bibfield  {title}
  {\bibinfo {title} {Ion-exchange reactions and physical properties of the mica
  analogue kniaso4},\ }\href@noop {} {\bibfield  {journal} {\bibinfo  {journal}
  {J. Solid State Chem.}\ }\textbf {\bibinfo {volume} {69}},\ \bibinfo {pages}
  {240} (\bibinfo {year} {1987})}\BibitemShut {NoStop}%
\bibitem [{\citenamefont {Buckley}\ \emph {et~al.}(1988)\citenamefont
  {Buckley}, \citenamefont {Bramwell}, \citenamefont {Day},\ and\ \citenamefont
  {Harrison}}]{Buckley1988}%
  \BibitemOpen
  \bibfield  {author} {\bibinfo {author} {\bibfnamefont {A.}~\bibnamefont
  {Buckley}}, \bibinfo {author} {\bibfnamefont {S.}~\bibnamefont {Bramwell}},
  \bibinfo {author} {\bibfnamefont {P.}~\bibnamefont {Day}},\ and\ \bibinfo
  {author} {\bibfnamefont {W.}~\bibnamefont {Harrison}},\ }\bibfield  {title}
  {\bibinfo {title} {{The crystal structure of potassium nickel arsenate;
  KNiAsO$_4$}},\ }\href@noop {} {\bibfield  {journal} {\bibinfo  {journal}
  {Zeitschrift f{\"u}r Naturforschung B}\ }\textbf {\bibinfo {volume} {43}},\
  \bibinfo {pages} {1053} (\bibinfo {year} {1988})}\BibitemShut {NoStop}%
\bibitem [{\citenamefont {Beneke}\ and\ \citenamefont
  {Lagaly}(1982)}]{Beneke1982}%
  \BibitemOpen
  \bibfield  {author} {\bibinfo {author} {\bibfnamefont {K.}~\bibnamefont
  {Beneke}}\ and\ \bibinfo {author} {\bibfnamefont {G.}~\bibnamefont
  {Lagaly}},\ }\bibfield  {title} {\bibinfo {title} {{The brittle mica-like
  KNiAsO$_4$ and its organic derivatives}},\ }\href@noop {} {\bibfield
  {journal} {\bibinfo  {journal} {Clay Miner.}\ }\textbf {\bibinfo {volume}
  {17}},\ \bibinfo {pages} {175} (\bibinfo {year} {1982})}\BibitemShut
  {NoStop}%
\bibitem [{\citenamefont {Bramwell}\ \emph {et~al.}(1988)\citenamefont
  {Bramwell}, \citenamefont {Buckley}, \citenamefont {Visser},\ and\
  \citenamefont {Day}}]{Bramwell1988}%
  \BibitemOpen
  \bibfield  {author} {\bibinfo {author} {\bibfnamefont {S.}~\bibnamefont
  {Bramwell}}, \bibinfo {author} {\bibfnamefont {A.}~\bibnamefont {Buckley}},
  \bibinfo {author} {\bibfnamefont {D.}~\bibnamefont {Visser}},\ and\ \bibinfo
  {author} {\bibfnamefont {P.}~\bibnamefont {Day}},\ }\bibfield  {title}
  {\bibinfo {title} {{Magnetic susceptibility study of KNiAsO$_4$,
  HMnAsO$_4${\textperiodcentered} H$_2$O and their organic-intercalated
  derivatives}},\ }\href@noop {} {\bibfield  {journal} {\bibinfo  {journal}
  {Phys. and Chem. Minerals}\ }\textbf {\bibinfo {volume} {15}},\ \bibinfo
  {pages} {465} (\bibinfo {year} {1988})}\BibitemShut {NoStop}%
\bibitem [{\citenamefont {Calder}\ \emph {et~al.}(2018)\citenamefont {Calder},
  \citenamefont {An}, \citenamefont {Boehler}, \citenamefont {Dela~Cruz},
  \citenamefont {Frontzek}, \citenamefont {Guthrie}, \citenamefont {Haberl},
  \citenamefont {Huq}, \citenamefont {Kimber}, \citenamefont {Liu} \emph
  {et~al.}}]{Calder2018}%
  \BibitemOpen
  \bibfield  {author} {\bibinfo {author} {\bibfnamefont {S.}~\bibnamefont
  {Calder}}, \bibinfo {author} {\bibfnamefont {K.}~\bibnamefont {An}}, \bibinfo
  {author} {\bibfnamefont {R.}~\bibnamefont {Boehler}}, \bibinfo {author}
  {\bibfnamefont {C.}~\bibnamefont {Dela~Cruz}}, \bibinfo {author}
  {\bibfnamefont {M.}~\bibnamefont {Frontzek}}, \bibinfo {author}
  {\bibfnamefont {M.}~\bibnamefont {Guthrie}}, \bibinfo {author} {\bibfnamefont
  {B.}~\bibnamefont {Haberl}}, \bibinfo {author} {\bibfnamefont
  {A.}~\bibnamefont {Huq}}, \bibinfo {author} {\bibfnamefont {S.~A.}\
  \bibnamefont {Kimber}}, \bibinfo {author} {\bibfnamefont {J.}~\bibnamefont
  {Liu}}, \emph {et~al.},\ }\bibfield  {title} {\bibinfo {title} {A suite-level
  review of the neutron powder diffraction instruments at oak ridge national
  laboratory},\ }\href@noop {} {\bibfield  {journal} {\bibinfo  {journal}
  {Review of Scientific Instruments}\ }\textbf {\bibinfo {volume} {89}},\
  \bibinfo {pages} {092701} (\bibinfo {year} {2018})}\BibitemShut {NoStop}%
\bibitem [{\citenamefont
  {Rodr{\'{\i}}guez-Carvajal}(1993)}]{Rodriguez-Carvajal1993}%
  \BibitemOpen
  \bibfield  {author} {\bibinfo {author} {\bibfnamefont {J.}~\bibnamefont
  {Rodr{\'{\i}}guez-Carvajal}},\ }\bibfield  {title} {\bibinfo {title} {{Recent
  advances in magnetic structure determination by neutron powder
  diffraction}},\ }\href {https://doi.org/10.1016/0921-4526(93)90108-I}
  {\bibfield  {journal} {\bibinfo  {journal} {Physica B: Condensed Matter}\
  }\textbf {\bibinfo {volume} {192}},\ \bibinfo {pages} {55} (\bibinfo {year}
  {1993})}\BibitemShut {NoStop}%
\bibitem [{\citenamefont {Wills}(2000)}]{Wills2000}%
  \BibitemOpen
  \bibfield  {author} {\bibinfo {author} {\bibfnamefont {A.}~\bibnamefont
  {Wills}},\ }\bibfield  {title} {\bibinfo {title} {{A new protocol for the
  determination of magnetic structures using simulated annealing and
  representational analysis (SARAh)}},\ }\href
  {https://doi.org/https://doi.org/10.1016/S0921-4526(99)01722-6} {\bibfield
  {journal} {\bibinfo  {journal} {Phys. B}\ }\textbf {\bibinfo {volume}
  {276-278}},\ \bibinfo {pages} {680 } (\bibinfo {year} {2000})}\BibitemShut
  {NoStop}%
\bibitem [{\citenamefont {Aroyo}\ \emph
  {et~al.}(2006{\natexlab{a}})\citenamefont {Aroyo}, \citenamefont
  {Perez-Mato}, \citenamefont {Capillas}, \citenamefont {Kroumova},
  \citenamefont {Ivantchev}, \citenamefont {Madariaga}, \citenamefont {Kirov},\
  and\ \citenamefont {Wondratschek}}]{Aroyo2006a}%
  \BibitemOpen
  \bibfield  {author} {\bibinfo {author} {\bibfnamefont {M.}~\bibnamefont
  {Aroyo}}, \bibinfo {author} {\bibfnamefont {J.}~\bibnamefont {Perez-Mato}},
  \bibinfo {author} {\bibfnamefont {C.}~\bibnamefont {Capillas}}, \bibinfo
  {author} {\bibfnamefont {E.}~\bibnamefont {Kroumova}}, \bibinfo {author}
  {\bibfnamefont {S.}~\bibnamefont {Ivantchev}}, \bibinfo {author}
  {\bibfnamefont {G.}~\bibnamefont {Madariaga}}, \bibinfo {author}
  {\bibfnamefont {A.}~\bibnamefont {Kirov}},\ and\ \bibinfo {author}
  {\bibfnamefont {H.}~\bibnamefont {Wondratschek}},\ }\bibfield  {title}
  {\bibinfo {title} {{Bilbao crystallographic server: I. Databases and
  crystallographic computing programs}},\ }\href
  {https://doi.org/10.1524/zkri.2006.221.1.15} {\bibfield  {journal} {\bibinfo
  {journal} {Z. KRIST.}\ }\textbf {\bibinfo {volume} {221}},\ \bibinfo {pages}
  {15} (\bibinfo {year} {2006}{\natexlab{a}})}\BibitemShut {NoStop}%
\bibitem [{\citenamefont {Aroyo}\ \emph
  {et~al.}(2006{\natexlab{b}})\citenamefont {Aroyo}, \citenamefont {Kirov},
  \citenamefont {Capillas}, \citenamefont {Perez-Mato},\ and\ \citenamefont
  {Wondratschek}}]{Aroyo2006b}%
  \BibitemOpen
  \bibfield  {author} {\bibinfo {author} {\bibfnamefont {M.}~\bibnamefont
  {Aroyo}}, \bibinfo {author} {\bibfnamefont {A.}~\bibnamefont {Kirov}},
  \bibinfo {author} {\bibfnamefont {C.}~\bibnamefont {Capillas}}, \bibinfo
  {author} {\bibfnamefont {J.}~\bibnamefont {Perez-Mato}},\ and\ \bibinfo
  {author} {\bibfnamefont {H.}~\bibnamefont {Wondratschek}},\ }\bibfield
  {title} {\bibinfo {title} {{Bilbao Crystallographic Server. II.
  Representations of crystallographic point groups and space groups}},\ }\href
  {https://doi.org/10.1107/S0108767305040286} {\bibfield  {journal} {\bibinfo
  {journal} {Acta Crystallogr. Sec. A}\ }\textbf {\bibinfo {volume} {62}},\
  \bibinfo {pages} {115} (\bibinfo {year} {2006}{\natexlab{b}})}\BibitemShut
  {NoStop}%
\bibitem [{\citenamefont {Aroyo}\ \emph {et~al.}(2011)\citenamefont {Aroyo},
  \citenamefont {Perez-Mato}, \citenamefont {Orobengoa}, \citenamefont {Tasci},
  \citenamefont {De~La~Flor},\ and\ \citenamefont {Kirov}}]{Aroyo2011}%
  \BibitemOpen
  \bibfield  {author} {\bibinfo {author} {\bibfnamefont {M.}~\bibnamefont
  {Aroyo}}, \bibinfo {author} {\bibfnamefont {J.}~\bibnamefont {Perez-Mato}},
  \bibinfo {author} {\bibfnamefont {D.}~\bibnamefont {Orobengoa}}, \bibinfo
  {author} {\bibfnamefont {E.}~\bibnamefont {Tasci}}, \bibinfo {author}
  {\bibfnamefont {G.}~\bibnamefont {De~La~Flor}},\ and\ \bibinfo {author}
  {\bibfnamefont {A.}~\bibnamefont {Kirov}},\ }\bibfield  {title} {\bibinfo
  {title} {{Crystallography online: Bilbao crystallographic server}},\ }\href
  {https://www2.scopus.com/inward/record.uri?eid=2-s2.0-80955140447&partnerID=40&md5=488772b9e21d2636a3952f66ae80ae84}
  {\bibfield  {journal} {\bibinfo  {journal} {Bulg. Chem. Commun.}\ }\textbf
  {\bibinfo {volume} {43}},\ \bibinfo {pages} {183} (\bibinfo {year}
  {2011})}\BibitemShut {NoStop}%
\bibitem [{\citenamefont {Momma}\ and\ \citenamefont
  {Izumi}(2011)}]{Momma2011}%
  \BibitemOpen
  \bibfield  {author} {\bibinfo {author} {\bibfnamefont {K.}~\bibnamefont
  {Momma}}\ and\ \bibinfo {author} {\bibfnamefont {F.}~\bibnamefont {Izumi}},\
  }\bibfield  {title} {\bibinfo {title} {{VESTA 3 for three dimensional
  visualization of crystal, volumetric and morphology data}},\ }\href@noop {}
  {\bibfield  {journal} {\bibinfo  {journal} {J. Appl. Crysatllogr.}\ }\textbf
  {\bibinfo {volume} {44}},\ \bibinfo {pages} {1272} (\bibinfo {year}
  {2011})}\BibitemShut {NoStop}%
\bibitem [{\citenamefont {Zaliznyak}\ \emph {et~al.}(2017)\citenamefont
  {Zaliznyak}, \citenamefont {Savici}, \citenamefont {Garlea}, \citenamefont
  {Winn}, \citenamefont {Filges}, \citenamefont {Schneeloch}, \citenamefont
  {Tranquada}, \citenamefont {Gu}, \citenamefont {Wang},\ and\ \citenamefont
  {Petrovic}}]{Zaliznyak2017}%
  \BibitemOpen
  \bibfield  {author} {\bibinfo {author} {\bibfnamefont {I.~A.}\ \bibnamefont
  {Zaliznyak}}, \bibinfo {author} {\bibfnamefont {A.~T.}\ \bibnamefont
  {Savici}}, \bibinfo {author} {\bibfnamefont {V.~O.}\ \bibnamefont {Garlea}},
  \bibinfo {author} {\bibfnamefont {B.}~\bibnamefont {Winn}}, \bibinfo {author}
  {\bibfnamefont {U.}~\bibnamefont {Filges}}, \bibinfo {author} {\bibfnamefont
  {J.}~\bibnamefont {Schneeloch}}, \bibinfo {author} {\bibfnamefont {J.~M.}\
  \bibnamefont {Tranquada}}, \bibinfo {author} {\bibfnamefont {G.}~\bibnamefont
  {Gu}}, \bibinfo {author} {\bibfnamefont {A.}~\bibnamefont {Wang}},\ and\
  \bibinfo {author} {\bibfnamefont {C.}~\bibnamefont {Petrovic}},\ }\bibfield
  {title} {\bibinfo {title} {{Polarized neutron scattering on {HYSPEC}: the
  {HYbrid} {SPECtrometer} at {SNS}}},\ }\href
  {https://doi.org/10.1088/1742-6596/862/1/012030} {\bibfield  {journal}
  {\bibinfo  {journal} {J. Phys.: Conf. Ser.}\ }\textbf {\bibinfo {volume}
  {862}},\ \bibinfo {pages} {012030} (\bibinfo {year} {2017})}\BibitemShut
  {NoStop}%
\bibitem [{\citenamefont {Samarakoon}\ \emph {et~al.}(2020)\citenamefont
  {Samarakoon}, \citenamefont {Barros}, \citenamefont {Li}, \citenamefont
  {Eisenbach}, \citenamefont {Zhang}, \citenamefont {Ye}, \citenamefont
  {Sharma}, \citenamefont {Dun}, \citenamefont {Zhou}, \citenamefont {Grigera},
  \citenamefont {Batista},\ and\ \citenamefont {Tennant}}]{Samarakoon2020}%
  \BibitemOpen
  \bibfield  {author} {\bibinfo {author} {\bibfnamefont {A.~M.}\ \bibnamefont
  {Samarakoon}}, \bibinfo {author} {\bibfnamefont {K.}~\bibnamefont {Barros}},
  \bibinfo {author} {\bibfnamefont {Y.~W.}\ \bibnamefont {Li}}, \bibinfo
  {author} {\bibfnamefont {M.}~\bibnamefont {Eisenbach}}, \bibinfo {author}
  {\bibfnamefont {Q.}~\bibnamefont {Zhang}}, \bibinfo {author} {\bibfnamefont
  {F.}~\bibnamefont {Ye}}, \bibinfo {author} {\bibfnamefont {V.}~\bibnamefont
  {Sharma}}, \bibinfo {author} {\bibfnamefont {Z.}~\bibnamefont {Dun}},
  \bibinfo {author} {\bibfnamefont {H.}~\bibnamefont {Zhou}}, \bibinfo {author}
  {\bibfnamefont {S.~A.}\ \bibnamefont {Grigera}}, \bibinfo {author}
  {\bibfnamefont {C.~D.}\ \bibnamefont {Batista}},\ and\ \bibinfo {author}
  {\bibfnamefont {D.~A.}\ \bibnamefont {Tennant}},\ }\bibfield  {title}
  {\bibinfo {title} {{Machine-learning-assisted insight into spin ice
  Dy$_2$Ti$_2$O$_7$}},\ }\href {https://doi.org/10.1038/s41467-020-14660-y}
  {\bibfield  {journal} {\bibinfo  {journal} {Nat. Comm.}\ }\textbf {\bibinfo
  {volume} {11}},\ \bibinfo {pages} {892} (\bibinfo {year} {2020})}\BibitemShut
  {NoStop}%
\bibitem [{\citenamefont {Toth}\ and\ \citenamefont {Lake}(2015)}]{Toth2015}%
  \BibitemOpen
  \bibfield  {author} {\bibinfo {author} {\bibfnamefont {S.}~\bibnamefont
  {Toth}}\ and\ \bibinfo {author} {\bibfnamefont {B.}~\bibnamefont {Lake}},\
  }\bibfield  {title} {\bibinfo {title} {{Linear spin wave theory for single-Q
  incommensurate magnetic structures}},\ }\href
  {https://doi.org/10.1088/0953-8984/27/26/166002} {\bibfield  {journal}
  {\bibinfo  {journal} {J. PHys.: Condens. Matter}\ }\textbf {\bibinfo {volume}
  {27}},\ \bibinfo {pages} {166002} (\bibinfo {year} {2015})}\BibitemShut
  {NoStop}%
\bibitem [{\citenamefont {Blaha}\ \emph {et~al.}(2001)\citenamefont {Blaha},
  \citenamefont {Schwarz}, \citenamefont {Madsen}, \citenamefont {Kvasnicka},\
  and\ \citenamefont {Luitz}}]{Blaha2001}%
  \BibitemOpen
  \bibfield  {author} {\bibinfo {author} {\bibfnamefont {P.}~\bibnamefont
  {Blaha}}, \bibinfo {author} {\bibfnamefont {K.}~\bibnamefont {Schwarz}},
  \bibinfo {author} {\bibfnamefont {G.}~\bibnamefont {Madsen}}, \bibinfo
  {author} {\bibfnamefont {D.}~\bibnamefont {Kvasnicka}},\ and\ \bibinfo
  {author} {\bibfnamefont {J.}~\bibnamefont {Luitz}},\ }\href@noop {} {\emph
  {\bibinfo {title} {WIEN2k, An augmented plane wave + lcal orbitals program
  for calculating crystal properties}}}\ (\bibinfo {year} {2001})\BibitemShut
  {NoStop}%
\bibitem [{\citenamefont {Perdew}\ \emph {et~al.}(1996)\citenamefont {Perdew},
  \citenamefont {Burke},\ and\ \citenamefont {Ernzerhof}}]{Perdew1996}%
  \BibitemOpen
  \bibfield  {author} {\bibinfo {author} {\bibfnamefont {J.~P.}\ \bibnamefont
  {Perdew}}, \bibinfo {author} {\bibfnamefont {K.}~\bibnamefont {Burke}},\ and\
  \bibinfo {author} {\bibfnamefont {M.}~\bibnamefont {Ernzerhof}},\ }\bibfield
  {title} {\bibinfo {title} {{Generalized Gradient Approximation Made
  Simple}},\ }\href {https://doi.org/10.1103/PhysRevLett.77.3865} {\bibfield
  {journal} {\bibinfo  {journal} {Phys. Rev. Lett.}\ }\textbf {\bibinfo
  {volume} {77}},\ \bibinfo {pages} {3865} (\bibinfo {year}
  {1996})}\BibitemShut {NoStop}%
\bibitem [{\citenamefont {Anisimov}\ \emph {et~al.}(1997)\citenamefont
  {Anisimov}, \citenamefont {Aryasetiawan},\ and\ \citenamefont
  {Lichtenstein}}]{Anisimov1997}%
  \BibitemOpen
  \bibfield  {author} {\bibinfo {author} {\bibfnamefont {V.~I.}\ \bibnamefont
  {Anisimov}}, \bibinfo {author} {\bibfnamefont {F.}~\bibnamefont
  {Aryasetiawan}},\ and\ \bibinfo {author} {\bibfnamefont {A.}~\bibnamefont
  {Lichtenstein}},\ }\bibfield  {title} {\bibinfo {title} {First-principles
  calculations of the electronic structure and spectra of strongly correlated
  systems: the lda+ u method},\ }\href@noop {} {\bibfield  {journal} {\bibinfo
  {journal} {J. Condens. Matter Phys.}\ }\textbf {\bibinfo {volume} {9}},\
  \bibinfo {pages} {767} (\bibinfo {year} {1997})}\BibitemShut {NoStop}%
\bibitem [{\citenamefont {Pokharel}\ \emph {et~al.}(2018)\citenamefont
  {Pokharel}, \citenamefont {May}, \citenamefont {Parker}, \citenamefont
  {Calder}, \citenamefont {Ehlers}, \citenamefont {Huq}, \citenamefont
  {Kimber}, \citenamefont {Arachchige}, \citenamefont {Poudel}, \citenamefont
  {McGuire} \emph {et~al.}}]{Pokharel2018}%
  \BibitemOpen
  \bibfield  {author} {\bibinfo {author} {\bibfnamefont {G.}~\bibnamefont
  {Pokharel}}, \bibinfo {author} {\bibfnamefont {A.}~\bibnamefont {May}},
  \bibinfo {author} {\bibfnamefont {D.}~\bibnamefont {Parker}}, \bibinfo
  {author} {\bibfnamefont {S.}~\bibnamefont {Calder}}, \bibinfo {author}
  {\bibfnamefont {G.}~\bibnamefont {Ehlers}}, \bibinfo {author} {\bibfnamefont
  {A.}~\bibnamefont {Huq}}, \bibinfo {author} {\bibfnamefont {S.}~\bibnamefont
  {Kimber}}, \bibinfo {author} {\bibfnamefont {H.~S.}\ \bibnamefont
  {Arachchige}}, \bibinfo {author} {\bibfnamefont {L.}~\bibnamefont {Poudel}},
  \bibinfo {author} {\bibfnamefont {M.}~\bibnamefont {McGuire}}, \emph
  {et~al.},\ }\bibfield  {title} {\bibinfo {title} {{Negative thermal expansion
  and magnetoelastic coupling in the breathing pyrochlore lattice material
  LiGaCr$_4$S$_8$}},\ }\href@noop {} {\bibfield  {journal} {\bibinfo  {journal}
  {Physical Review B}\ }\textbf {\bibinfo {volume} {97}},\ \bibinfo {pages}
  {134117} (\bibinfo {year} {2018})}\BibitemShut {NoStop}%
\bibitem [{\citenamefont {Niedziela}\ \emph {et~al.}(2021)\citenamefont
  {Niedziela}, \citenamefont {Sanjeewa}, \citenamefont {Podlesnyak},
  \citenamefont {DeBeer-Schmitt}, \citenamefont {Kuhn}, \citenamefont {de~la
  Cruz}, \citenamefont {Parker}, \citenamefont {Page},\ and\ \citenamefont
  {Sefat}}]{Niedziela2021}%
  \BibitemOpen
  \bibfield  {author} {\bibinfo {author} {\bibfnamefont {J.~L.}\ \bibnamefont
  {Niedziela}}, \bibinfo {author} {\bibfnamefont {L.~D.}\ \bibnamefont
  {Sanjeewa}}, \bibinfo {author} {\bibfnamefont {A.~A.}\ \bibnamefont
  {Podlesnyak}}, \bibinfo {author} {\bibfnamefont {L.}~\bibnamefont
  {DeBeer-Schmitt}}, \bibinfo {author} {\bibfnamefont {S.~J.}\ \bibnamefont
  {Kuhn}}, \bibinfo {author} {\bibfnamefont {C.}~\bibnamefont {de~la Cruz}},
  \bibinfo {author} {\bibfnamefont {D.~S.}\ \bibnamefont {Parker}}, \bibinfo
  {author} {\bibfnamefont {K.}~\bibnamefont {Page}},\ and\ \bibinfo {author}
  {\bibfnamefont {A.~S.}\ \bibnamefont {Sefat}},\ }\bibfield  {title} {\bibinfo
  {title} {{Magnetoelastic coupling, negative thermal expansion, and
  two-dimensional magnetic excitations in FeAs}},\ }\href@noop {} {\bibfield
  {journal} {\bibinfo  {journal} {Phys. Rev. B}\ }\textbf {\bibinfo {volume}
  {103}},\ \bibinfo {pages} {094431} (\bibinfo {year} {2021})}\BibitemShut
  {NoStop}%
\bibitem [{\citenamefont {Sanjeewa}\ \emph {et~al.}(2020)\citenamefont
  {Sanjeewa}, \citenamefont {Xing}, \citenamefont {Taddei}, \citenamefont
  {Parker}, \citenamefont {Custelcean}, \citenamefont {dela Cruz},\ and\
  \citenamefont {Sefat}}]{Sanjeewa2020}%
  \BibitemOpen
  \bibfield  {author} {\bibinfo {author} {\bibfnamefont {L.~D.}\ \bibnamefont
  {Sanjeewa}}, \bibinfo {author} {\bibfnamefont {J.}~\bibnamefont {Xing}},
  \bibinfo {author} {\bibfnamefont {K.~M.}\ \bibnamefont {Taddei}}, \bibinfo
  {author} {\bibfnamefont {D.}~\bibnamefont {Parker}}, \bibinfo {author}
  {\bibfnamefont {R.}~\bibnamefont {Custelcean}}, \bibinfo {author}
  {\bibfnamefont {C.}~\bibnamefont {dela Cruz}},\ and\ \bibinfo {author}
  {\bibfnamefont {A.~S.}\ \bibnamefont {Sefat}},\ }\bibfield  {title} {\bibinfo
  {title} {{Evidence of Ba-substitution induced spin-canting in the magnetic
  Weyl semimetal $\mathrm{Eu}{\mathrm{Cd}}_{2}{\mathrm{As}}_{2}$}},\ }\href
  {https://doi.org/10.1103/PhysRevB.102.104404} {\bibfield  {journal} {\bibinfo
   {journal} {Phys. Rev. B}\ }\textbf {\bibinfo {volume} {102}},\ \bibinfo
  {pages} {104404} (\bibinfo {year} {2020})}\BibitemShut {NoStop}%
\bibitem [{\citenamefont {May}\ \emph {et~al.}(2012)\citenamefont {May},
  \citenamefont {McGuire}, \citenamefont {Cao}, \citenamefont {Sergueev},
  \citenamefont {Cantoni}, \citenamefont {Chakoumakos}, \citenamefont
  {Parker},\ and\ \citenamefont {Sales}}]{May2012}%
  \BibitemOpen
  \bibfield  {author} {\bibinfo {author} {\bibfnamefont {A.~F.}\ \bibnamefont
  {May}}, \bibinfo {author} {\bibfnamefont {M.~A.}\ \bibnamefont {McGuire}},
  \bibinfo {author} {\bibfnamefont {H.}~\bibnamefont {Cao}}, \bibinfo {author}
  {\bibfnamefont {I.}~\bibnamefont {Sergueev}}, \bibinfo {author}
  {\bibfnamefont {C.}~\bibnamefont {Cantoni}}, \bibinfo {author} {\bibfnamefont
  {B.~C.}\ \bibnamefont {Chakoumakos}}, \bibinfo {author} {\bibfnamefont
  {D.~S.}\ \bibnamefont {Parker}},\ and\ \bibinfo {author} {\bibfnamefont
  {B.~C.}\ \bibnamefont {Sales}},\ }\bibfield  {title} {\bibinfo {title}
  {{SSpin reorientation in TlFe$_{1.6}$Se$_2$ with complete vacancy
  ordering}},\ }\href {https://doi.org/10.1103/PhysRevLett.109.077003}
  {\bibfield  {journal} {\bibinfo  {journal} {Physical Review Letters}\
  }\textbf {\bibinfo {volume} {109}},\ \bibinfo {pages} {077003} (\bibinfo
  {year} {2012})}\BibitemShut {NoStop}%
\bibitem [{\citenamefont {Shanavas}\ \emph {et~al.}(2014)\citenamefont
  {Shanavas}, \citenamefont {Parker},\ and\ \citenamefont
  {Singh}}]{Shanavas2014}%
  \BibitemOpen
  \bibfield  {author} {\bibinfo {author} {\bibfnamefont {K.}~\bibnamefont
  {Shanavas}}, \bibinfo {author} {\bibfnamefont {D.}~\bibnamefont {Parker}},\
  and\ \bibinfo {author} {\bibfnamefont {D.~J.}\ \bibnamefont {Singh}},\
  }\bibfield  {title} {\bibinfo {title} {{Theoretical study on the role of
  dynamics on the unusual magnetic properties in MnBi}},\ }\href@noop {}
  {\bibfield  {journal} {\bibinfo  {journal} {Sci. Rep.}\ }\textbf {\bibinfo
  {volume} {4}},\ \bibinfo {pages} {1} (\bibinfo {year} {2014})}\BibitemShut
  {NoStop}%
\bibitem [{\citenamefont {Bramwell}\ \emph {et~al.}(1994)\citenamefont
  {Bramwell}, \citenamefont {Buckley},\ and\ \citenamefont
  {Day}}]{Bramwell1994}%
  \BibitemOpen
  \bibfield  {author} {\bibinfo {author} {\bibfnamefont {S.~T.}\ \bibnamefont
  {Bramwell}}, \bibinfo {author} {\bibfnamefont {A.~M.}\ \bibnamefont
  {Buckley}},\ and\ \bibinfo {author} {\bibfnamefont {P.}~\bibnamefont {Day}},\
  }\bibfield  {title} {\bibinfo {title} {The magnetic structure of kniaso$_4$:
  a two-dimensional honeycomb lattice},\ }\href@noop {} {\bibfield  {journal}
  {\bibinfo  {journal} {J. Solid State Chem.}\ }\textbf {\bibinfo {volume}
  {111}},\ \bibinfo {pages} {48} (\bibinfo {year} {1994})}\BibitemShut
  {NoStop}%
\bibitem [{\citenamefont {Hiroi}\ \emph {et~al.}(2001)\citenamefont {Hiroi},
  \citenamefont {Hanawa}, \citenamefont {Kobayashi}, \citenamefont {Nohara},
  \citenamefont {Takagi}, \citenamefont {Kato},\ and\ \citenamefont
  {Takigawa}}]{Hiroi2001}%
  \BibitemOpen
  \bibfield  {author} {\bibinfo {author} {\bibfnamefont {Z.}~\bibnamefont
  {Hiroi}}, \bibinfo {author} {\bibfnamefont {M.}~\bibnamefont {Hanawa}},
  \bibinfo {author} {\bibfnamefont {N.}~\bibnamefont {Kobayashi}}, \bibinfo
  {author} {\bibfnamefont {M.}~\bibnamefont {Nohara}}, \bibinfo {author}
  {\bibfnamefont {H.}~\bibnamefont {Takagi}}, \bibinfo {author} {\bibfnamefont
  {Y.}~\bibnamefont {Kato}},\ and\ \bibinfo {author} {\bibfnamefont
  {M.}~\bibnamefont {Takigawa}},\ }\bibfield  {title} {\bibinfo {title}
  {Spin-1/2 kagomé-like lattice in volborthite
  Cu$_3$V$_2$O$_7$(OH)$_2$·2H$_2$O},\ }\href
  {https://doi.org/10.1143/JPSJ.70.3377} {\bibfield  {journal} {\bibinfo
  {journal} {J. Phys. Soc. Japan}\ }\textbf {\bibinfo {volume} {70}},\ \bibinfo
  {pages} {3377} (\bibinfo {year} {2001})},\ \Eprint
  {https://arxiv.org/abs/https://doi.org/10.1143/JPSJ.70.3377}
  {https://doi.org/10.1143/JPSJ.70.3377} \BibitemShut {NoStop}%
\bibitem [{\citenamefont {Vasiliev}\ \emph {et~al.}(2018)\citenamefont
  {Vasiliev}, \citenamefont {Volkova}, \citenamefont {Zvereva},\ and\
  \citenamefont {Markina}}]{Vasiliev2018}%
  \BibitemOpen
  \bibfield  {author} {\bibinfo {author} {\bibfnamefont {A.}~\bibnamefont
  {Vasiliev}}, \bibinfo {author} {\bibfnamefont {O.}~\bibnamefont {Volkova}},
  \bibinfo {author} {\bibfnamefont {E.}~\bibnamefont {Zvereva}},\ and\ \bibinfo
  {author} {\bibfnamefont {M.}~\bibnamefont {Markina}},\ }\bibfield  {title}
  {\bibinfo {title} {Milestones of low-d quantum magnetism},\ }\href@noop {}
  {\bibfield  {journal} {\bibinfo  {journal} {npj Quantum Mater}\ }\textbf
  {\bibinfo {volume} {3}},\ \bibinfo {pages} {1} (\bibinfo {year}
  {2018})}\BibitemShut {NoStop}%
\bibitem [{\citenamefont {Taddei}\ \emph {et~al.}(2022)\citenamefont {Taddei},
  \citenamefont {Yin}, \citenamefont {Sanjeewa}, \citenamefont {Li},
  \citenamefont {Xing}, \citenamefont {dela Cruz}, \citenamefont {Phelan},
  \citenamefont {Sefat},\ and\ \citenamefont {Parker}}]{Taddei2022}%
  \BibitemOpen
  \bibfield  {author} {\bibinfo {author} {\bibfnamefont {K.}~\bibnamefont
  {Taddei}}, \bibinfo {author} {\bibfnamefont {L.}~\bibnamefont {Yin}},
  \bibinfo {author} {\bibfnamefont {L.}~\bibnamefont {Sanjeewa}}, \bibinfo
  {author} {\bibfnamefont {Y.}~\bibnamefont {Li}}, \bibinfo {author}
  {\bibfnamefont {J.}~\bibnamefont {Xing}}, \bibinfo {author} {\bibfnamefont
  {C.}~\bibnamefont {dela Cruz}}, \bibinfo {author} {\bibfnamefont
  {D.}~\bibnamefont {Phelan}}, \bibinfo {author} {\bibfnamefont
  {A.}~\bibnamefont {Sefat}},\ and\ \bibinfo {author} {\bibfnamefont
  {D.}~\bibnamefont {Parker}},\ }\bibfield  {title} {\bibinfo {title} {{Single
  pair of Weyl nodes in the spin-canted structure of EuCd$_2$As$_2$}},\
  }\href@noop {} {\bibfield  {journal} {\bibinfo  {journal} {Phys. Rev. B}\
  }\textbf {\bibinfo {volume} {105}},\ \bibinfo {pages} {L140401} (\bibinfo
  {year} {2022})}\BibitemShut {NoStop}%
\bibitem [{\citenamefont {Taddei}\ \emph {et~al.}(2019)\citenamefont {Taddei},
  \citenamefont {Sanjeewa}, \citenamefont {Kolis}, \citenamefont {Sefat},
  \citenamefont {De~La~Cruz},\ and\ \citenamefont {Pajerowski}}]{Taddei2019}%
  \BibitemOpen
  \bibfield  {author} {\bibinfo {author} {\bibfnamefont {K.~M.}\ \bibnamefont
  {Taddei}}, \bibinfo {author} {\bibfnamefont {L.}~\bibnamefont {Sanjeewa}},
  \bibinfo {author} {\bibfnamefont {J.~W.}\ \bibnamefont {Kolis}}, \bibinfo
  {author} {\bibfnamefont {A.~S.}\ \bibnamefont {Sefat}}, \bibinfo {author}
  {\bibfnamefont {C.}~\bibnamefont {De~La~Cruz}},\ and\ \bibinfo {author}
  {\bibfnamefont {D.~M.}\ \bibnamefont {Pajerowski}},\ }\bibfield  {title}
  {\bibinfo {title} {{Local-Ising-type magnetic order and metamagnetism in the
  rare-earth pyrogermanate Er$_2$Ge$_2$O$_7$}},\ }\href@noop {} {\bibfield
  {journal} {\bibinfo  {journal} {Phys. Rev. Mater.}\ }\textbf {\bibinfo
  {volume} {3}},\ \bibinfo {pages} {014405} (\bibinfo {year}
  {2019})}\BibitemShut {NoStop}%
\bibitem [{\citenamefont {Xing}\ \emph {et~al.}(2021)\citenamefont {Xing},
  \citenamefont {Taddei}, \citenamefont {Sanjeewa}, \citenamefont {Fishman},
  \citenamefont {Daum}, \citenamefont {Mourigal}, \citenamefont {dela Cruz},\
  and\ \citenamefont {Sefat}}]{Xing2021}%
  \BibitemOpen
  \bibfield  {author} {\bibinfo {author} {\bibfnamefont {J.}~\bibnamefont
  {Xing}}, \bibinfo {author} {\bibfnamefont {K.~M.}\ \bibnamefont {Taddei}},
  \bibinfo {author} {\bibfnamefont {L.~D.}\ \bibnamefont {Sanjeewa}}, \bibinfo
  {author} {\bibfnamefont {R.~S.}\ \bibnamefont {Fishman}}, \bibinfo {author}
  {\bibfnamefont {M.}~\bibnamefont {Daum}}, \bibinfo {author} {\bibfnamefont
  {M.}~\bibnamefont {Mourigal}}, \bibinfo {author} {\bibfnamefont
  {C.}~\bibnamefont {dela Cruz}},\ and\ \bibinfo {author} {\bibfnamefont
  {A.~S.}\ \bibnamefont {Sefat}},\ }\bibfield  {title} {\bibinfo {title}
  {{Stripe antiferromagnetic ground state of the ideal triangular lattice
  compound ${\mathrm{KErSe}}_{2}$}},\ }\href
  {https://doi.org/10.1103/PhysRevB.103.144413} {\bibfield  {journal} {\bibinfo
   {journal} {Phys. Rev. B}\ }\textbf {\bibinfo {volume} {103}},\ \bibinfo
  {pages} {144413} (\bibinfo {year} {2021})}\BibitemShut {NoStop}%
\bibitem [{\citenamefont {Stavrou}\ \emph {et~al.}(2015)\citenamefont
  {Stavrou}, \citenamefont {Chen}, \citenamefont {Oganov}, \citenamefont
  {Wang}, \citenamefont {Yan}, \citenamefont {Luo}, \citenamefont {Chen},\ and\
  \citenamefont {Goncharov}}]{Stavrou2015}%
  \BibitemOpen
  \bibfield  {author} {\bibinfo {author} {\bibfnamefont {E.}~\bibnamefont
  {Stavrou}}, \bibinfo {author} {\bibfnamefont {X.-J.}\ \bibnamefont {Chen}},
  \bibinfo {author} {\bibfnamefont {A.~R.}\ \bibnamefont {Oganov}}, \bibinfo
  {author} {\bibfnamefont {a.~F.}\ \bibnamefont {Wang}}, \bibinfo {author}
  {\bibfnamefont {Y.~J.}\ \bibnamefont {Yan}}, \bibinfo {author} {\bibfnamefont
  {X.~G.}\ \bibnamefont {Luo}}, \bibinfo {author} {\bibfnamefont {X.~H.}\
  \bibnamefont {Chen}},\ and\ \bibinfo {author} {\bibfnamefont {A.~F.}\
  \bibnamefont {Goncharov}},\ }\bibfield  {title} {\bibinfo {title} {{Formation
  of As-As Interlayer Bonding in the collapsed tetragonal phase of NaFe2As2
  under pressure}},\ }\href {https://doi.org/10.1038/srep09868} {\bibfield
  {journal} {\bibinfo  {journal} {Scientific Reports}\ }\textbf {\bibinfo
  {volume} {5}},\ \bibinfo {pages} {9868} (\bibinfo {year} {2015})},\ \Eprint
  {https://arxiv.org/abs/1505.05474} {arXiv:1505.05474} \BibitemShut {NoStop}%
\bibitem [{\citenamefont {Chaloupka}\ \emph {et~al.}(2013)\citenamefont
  {Chaloupka}, \citenamefont {Jackeli},\ and\ \citenamefont
  {Khaliullin}}]{Chaloupka2013}%
  \BibitemOpen
  \bibfield  {author} {\bibinfo {author} {\bibfnamefont {J.~C.~V.}\
  \bibnamefont {Chaloupka}}, \bibinfo {author} {\bibfnamefont {G.}~\bibnamefont
  {Jackeli}},\ and\ \bibinfo {author} {\bibfnamefont {G.}~\bibnamefont
  {Khaliullin}},\ }\bibfield  {title} {\bibinfo {title} {Zigzag magnetic order
  in the iridium oxide ${\mathrm{Na}}_{2}{\mathrm{IrO}}_{3}$},\ }\href
  {https://doi.org/10.1103/PhysRevLett.110.097204} {\bibfield  {journal}
  {\bibinfo  {journal} {Phys. Rev. Lett.}\ }\textbf {\bibinfo {volume} {110}},\
  \bibinfo {pages} {097204} (\bibinfo {year} {2013})}\BibitemShut {NoStop}%
\bibitem [{\citenamefont {Choi}\ \emph {et~al.}(2012)\citenamefont {Choi},
  \citenamefont {Coldea}, \citenamefont {Kolmogorov}, \citenamefont
  {Lancaster}, \citenamefont {Mazin}, \citenamefont {Blundell}, \citenamefont
  {Radaelli}, \citenamefont {Singh}, \citenamefont {Gegenwart}, \citenamefont
  {Choi}, \citenamefont {Cheong}, \citenamefont {Baker}, \citenamefont
  {Stock},\ and\ \citenamefont {Taylor}}]{Choi2012}%
  \BibitemOpen
  \bibfield  {author} {\bibinfo {author} {\bibfnamefont {S.~K.}\ \bibnamefont
  {Choi}}, \bibinfo {author} {\bibfnamefont {R.}~\bibnamefont {Coldea}},
  \bibinfo {author} {\bibfnamefont {A.~N.}\ \bibnamefont {Kolmogorov}},
  \bibinfo {author} {\bibfnamefont {T.}~\bibnamefont {Lancaster}}, \bibinfo
  {author} {\bibfnamefont {I.~I.}\ \bibnamefont {Mazin}}, \bibinfo {author}
  {\bibfnamefont {S.~J.}\ \bibnamefont {Blundell}}, \bibinfo {author}
  {\bibfnamefont {P.~G.}\ \bibnamefont {Radaelli}}, \bibinfo {author}
  {\bibfnamefont {Y.}~\bibnamefont {Singh}}, \bibinfo {author} {\bibfnamefont
  {P.}~\bibnamefont {Gegenwart}}, \bibinfo {author} {\bibfnamefont {K.~R.}\
  \bibnamefont {Choi}}, \bibinfo {author} {\bibfnamefont {S.-W.}\ \bibnamefont
  {Cheong}}, \bibinfo {author} {\bibfnamefont {P.~J.}\ \bibnamefont {Baker}},
  \bibinfo {author} {\bibfnamefont {C.}~\bibnamefont {Stock}},\ and\ \bibinfo
  {author} {\bibfnamefont {J.}~\bibnamefont {Taylor}},\ }\bibfield  {title}
  {\bibinfo {title} {{Spin Waves and Revised Crystal Structure of Honeycomb
  Iridate ${\mathrm{Na}}_{2}{\mathrm{IrO}}_{3}$}},\ }\href
  {https://doi.org/10.1103/PhysRevLett.108.127204} {\bibfield  {journal}
  {\bibinfo  {journal} {Phys. Rev. Lett.}\ }\textbf {\bibinfo {volume} {108}},\
  \bibinfo {pages} {127204} (\bibinfo {year} {2012})}\BibitemShut {NoStop}%
\bibitem [{\citenamefont {Halloran}\ \emph {et~al.}(2022)\citenamefont
  {Halloran}, \citenamefont {Desrochers}, \citenamefont {Zhang}, \citenamefont
  {Chen}, \citenamefont {Chern}, \citenamefont {Xu}, \citenamefont {Winn},
  \citenamefont {Graves-Brook}, \citenamefont {Stone}, \citenamefont
  {Kolesnikov} \emph {et~al.}}]{Halloran2022}%
  \BibitemOpen
  \bibfield  {author} {\bibinfo {author} {\bibfnamefont {T.}~\bibnamefont
  {Halloran}}, \bibinfo {author} {\bibfnamefont {F.}~\bibnamefont
  {Desrochers}}, \bibinfo {author} {\bibfnamefont {E.~Z.}\ \bibnamefont
  {Zhang}}, \bibinfo {author} {\bibfnamefont {T.}~\bibnamefont {Chen}},
  \bibinfo {author} {\bibfnamefont {L.~E.}\ \bibnamefont {Chern}}, \bibinfo
  {author} {\bibfnamefont {Z.}~\bibnamefont {Xu}}, \bibinfo {author}
  {\bibfnamefont {B.}~\bibnamefont {Winn}}, \bibinfo {author} {\bibfnamefont
  {M.}~\bibnamefont {Graves-Brook}}, \bibinfo {author} {\bibfnamefont
  {M.}~\bibnamefont {Stone}}, \bibinfo {author} {\bibfnamefont {A.~I.}\
  \bibnamefont {Kolesnikov}}, \emph {et~al.},\ }\bibfield  {title} {\bibinfo
  {title} {Geometrical frustration versus kitaev interactions in baco $ \_2
  $(aso $ \_4 $) $ \_2$},\ }\href@noop {} {\bibfield  {journal} {\bibinfo
  {journal} {arXiv preprint arXiv:2205.15262}\ } (\bibinfo {year}
  {2022})}\BibitemShut {NoStop}%
\bibitem [{\citenamefont {Rau}\ \emph {et~al.}(2014)\citenamefont {Rau},
  \citenamefont {Lee},\ and\ \citenamefont {Kee}}]{Rau2014}%
  \BibitemOpen
  \bibfield  {author} {\bibinfo {author} {\bibfnamefont {J.~G.}\ \bibnamefont
  {Rau}}, \bibinfo {author} {\bibfnamefont {E.~K.-H.}\ \bibnamefont {Lee}},\
  and\ \bibinfo {author} {\bibfnamefont {H.-Y.}\ \bibnamefont {Kee}},\
  }\bibfield  {title} {\bibinfo {title} {Generic spin model for the honeycomb
  iridates beyond the kitaev limit},\ }\href
  {https://doi.org/10.1103/PhysRevLett.112.077204} {\bibfield  {journal}
  {\bibinfo  {journal} {Phys. Rev. Lett.}\ }\textbf {\bibinfo {volume} {112}},\
  \bibinfo {pages} {077204} (\bibinfo {year} {2014})}\BibitemShut {NoStop}%
\bibitem [{\citenamefont {Matsuda}\ \emph {et~al.}(2019)\citenamefont
  {Matsuda}, \citenamefont {Dissanayake}, \citenamefont {Abernathy},
  \citenamefont {Qiu}, \citenamefont {Copley}, \citenamefont {Kumada},\ and\
  \citenamefont {Azuma}}]{Matsuda2019}%
  \BibitemOpen
  \bibfield  {author} {\bibinfo {author} {\bibfnamefont {M.}~\bibnamefont
  {Matsuda}}, \bibinfo {author} {\bibfnamefont {S.~E.}\ \bibnamefont
  {Dissanayake}}, \bibinfo {author} {\bibfnamefont {D.~L.}\ \bibnamefont
  {Abernathy}}, \bibinfo {author} {\bibfnamefont {Y.}~\bibnamefont {Qiu}},
  \bibinfo {author} {\bibfnamefont {J.~R.~D.}\ \bibnamefont {Copley}}, \bibinfo
  {author} {\bibfnamefont {N.}~\bibnamefont {Kumada}},\ and\ \bibinfo {author}
  {\bibfnamefont {M.}~\bibnamefont {Azuma}},\ }\bibfield  {title} {\bibinfo
  {title} {{Frustrated magnetic interactions in an $S=3/2$ bilayer honeycomb
  lattice compound
  ${\mathrm{Bi}}_{3}{\mathrm{Mn}}_{4}{\mathrm{O}}_{12}$(${\mathrm{NO}}_{3}$)}},\
  }\href {https://doi.org/10.1103/PhysRevB.100.134430} {\bibfield  {journal}
  {\bibinfo  {journal} {Phys. Rev. B}\ }\textbf {\bibinfo {volume} {100}},\
  \bibinfo {pages} {134430} (\bibinfo {year} {2019})}\BibitemShut {NoStop}%
\bibitem [{\citenamefont {Gao}\ \emph {et~al.}(2021)\citenamefont {Gao},
  \citenamefont {Chen}, \citenamefont {Wang}, \citenamefont {Chen},
  \citenamefont {Zhong}, \citenamefont {Abernathy}, \citenamefont {Xiao},\ and\
  \citenamefont {Dai}}]{Gao2021}%
  \BibitemOpen
  \bibfield  {author} {\bibinfo {author} {\bibfnamefont {B.}~\bibnamefont
  {Gao}}, \bibinfo {author} {\bibfnamefont {T.}~\bibnamefont {Chen}}, \bibinfo
  {author} {\bibfnamefont {C.}~\bibnamefont {Wang}}, \bibinfo {author}
  {\bibfnamefont {L.}~\bibnamefont {Chen}}, \bibinfo {author} {\bibfnamefont
  {R.}~\bibnamefont {Zhong}}, \bibinfo {author} {\bibfnamefont {D.~L.}\
  \bibnamefont {Abernathy}}, \bibinfo {author} {\bibfnamefont {D.}~\bibnamefont
  {Xiao}},\ and\ \bibinfo {author} {\bibfnamefont {P.}~\bibnamefont {Dai}},\
  }\bibfield  {title} {\bibinfo {title} {{Spin waves and Dirac magnons in a
  honeycomb-lattice zigzag antiferromagnet
  ${\text{BaNi}}_{2}{({\text{AsO}}_{4})}_{2}$}},\ }\href
  {https://doi.org/10.1103/PhysRevB.104.214432} {\bibfield  {journal} {\bibinfo
   {journal} {Phys. Rev. B}\ }\textbf {\bibinfo {volume} {104}},\ \bibinfo
  {pages} {214432} (\bibinfo {year} {2021})}\BibitemShut {NoStop}%
\bibitem [{\citenamefont {Chen}\ \emph {et~al.}(2019)\citenamefont {Chen},
  \citenamefont {Wang}, \citenamefont {Yan}, \citenamefont {Parker},
  \citenamefont {Zhou}, \citenamefont {Uwatoko},\ and\ \citenamefont
  {Cheng}}]{Chen2019}%
  \BibitemOpen
  \bibfield  {author} {\bibinfo {author} {\bibfnamefont {K.}~\bibnamefont
  {Chen}}, \bibinfo {author} {\bibfnamefont {B.}~\bibnamefont {Wang}}, \bibinfo
  {author} {\bibfnamefont {J.-Q.}\ \bibnamefont {Yan}}, \bibinfo {author}
  {\bibfnamefont {D.}~\bibnamefont {Parker}}, \bibinfo {author} {\bibfnamefont
  {J.-S.}\ \bibnamefont {Zhou}}, \bibinfo {author} {\bibfnamefont
  {Y.}~\bibnamefont {Uwatoko}},\ and\ \bibinfo {author} {\bibfnamefont {J.-G.}\
  \bibnamefont {Cheng}},\ }\bibfield  {title} {\bibinfo {title} {{Suppression
  of the antiferromagnetic metallic state in the pressurized MnBi$_2$Te$_4$
  single crystal}},\ }\href@noop {} {\bibfield  {journal} {\bibinfo  {journal}
  {Physical Review Materials}\ }\textbf {\bibinfo {volume} {3}},\ \bibinfo
  {pages} {094201} (\bibinfo {year} {2019})}\BibitemShut {NoStop}%
\bibitem [{\citenamefont {Yan}\ \emph {et~al.}(2020)\citenamefont {Yan},
  \citenamefont {Liu}, \citenamefont {Parker}, \citenamefont {Wu},
  \citenamefont {Aczel}, \citenamefont {Matsuda}, \citenamefont {McGuire},\
  and\ \citenamefont {Sales}}]{Yan2020}%
  \BibitemOpen
  \bibfield  {author} {\bibinfo {author} {\bibfnamefont {J.-Q.}\ \bibnamefont
  {Yan}}, \bibinfo {author} {\bibfnamefont {Y.}~\bibnamefont {Liu}}, \bibinfo
  {author} {\bibfnamefont {D.}~\bibnamefont {Parker}}, \bibinfo {author}
  {\bibfnamefont {Y.}~\bibnamefont {Wu}}, \bibinfo {author} {\bibfnamefont
  {A.}~\bibnamefont {Aczel}}, \bibinfo {author} {\bibfnamefont
  {M.}~\bibnamefont {Matsuda}}, \bibinfo {author} {\bibfnamefont
  {M.}~\bibnamefont {McGuire}},\ and\ \bibinfo {author} {\bibfnamefont
  {B.}~\bibnamefont {Sales}},\ }\bibfield  {title} {\bibinfo {title} {{A-type
  antiferromagnetic order in MnBi$_4$Te$_7$ and MnBi$_6$Te$_10$ single
  crystals}},\ }\href@noop {} {\bibfield  {journal} {\bibinfo  {journal} {Phys.
  Rev. Matt.}\ }\textbf {\bibinfo {volume} {4}},\ \bibinfo {pages} {054202}
  (\bibinfo {year} {2020})}\BibitemShut {NoStop}%
\bibitem [{\citenamefont {Williams}\ \emph {et~al.}(2016)\citenamefont
  {Williams}, \citenamefont {Taylor}, \citenamefont {Christianson},
  \citenamefont {Hahn}, \citenamefont {Fishman}, \citenamefont {Parker},
  \citenamefont {McGuire}, \citenamefont {Sales},\ and\ \citenamefont
  {Lumsden}}]{Williams2016}%
  \BibitemOpen
  \bibfield  {author} {\bibinfo {author} {\bibfnamefont {T.}~\bibnamefont
  {Williams}}, \bibinfo {author} {\bibfnamefont {A.}~\bibnamefont {Taylor}},
  \bibinfo {author} {\bibfnamefont {A.}~\bibnamefont {Christianson}}, \bibinfo
  {author} {\bibfnamefont {S.}~\bibnamefont {Hahn}}, \bibinfo {author}
  {\bibfnamefont {R.}~\bibnamefont {Fishman}}, \bibinfo {author} {\bibfnamefont
  {D.}~\bibnamefont {Parker}}, \bibinfo {author} {\bibfnamefont
  {M.}~\bibnamefont {McGuire}}, \bibinfo {author} {\bibfnamefont
  {B.}~\bibnamefont {Sales}},\ and\ \bibinfo {author} {\bibfnamefont
  {M.}~\bibnamefont {Lumsden}},\ }\bibfield  {title} {\bibinfo {title}
  {Extended magnetic exchange interactions in the high-temperature ferromagnet
  mnbi},\ }\href@noop {} {\bibfield  {journal} {\bibinfo  {journal} {Appl.
  Phys. Lett.}\ }\textbf {\bibinfo {volume} {108}},\ \bibinfo {pages} {192403}
  (\bibinfo {year} {2016})}\BibitemShut {NoStop}%
\bibitem [{\citenamefont {Lamichhane}\ \emph {et~al.}(2016)\citenamefont
  {Lamichhane}, \citenamefont {Taufour}, \citenamefont {Masters}, \citenamefont
  {Parker}, \citenamefont {Kaluarachchi}, \citenamefont {Thimmaiah},
  \citenamefont {Bud'ko},\ and\ \citenamefont {Canfield}}]{Lamichhane2016}%
  \BibitemOpen
  \bibfield  {author} {\bibinfo {author} {\bibfnamefont {T.~N.}\ \bibnamefont
  {Lamichhane}}, \bibinfo {author} {\bibfnamefont {V.}~\bibnamefont {Taufour}},
  \bibinfo {author} {\bibfnamefont {M.~W.}\ \bibnamefont {Masters}}, \bibinfo
  {author} {\bibfnamefont {D.~S.}\ \bibnamefont {Parker}}, \bibinfo {author}
  {\bibfnamefont {U.~S.}\ \bibnamefont {Kaluarachchi}}, \bibinfo {author}
  {\bibfnamefont {S.}~\bibnamefont {Thimmaiah}}, \bibinfo {author}
  {\bibfnamefont {S.~L.}\ \bibnamefont {Bud'ko}},\ and\ \bibinfo {author}
  {\bibfnamefont {P.~C.}\ \bibnamefont {Canfield}},\ }\bibfield  {title}
  {\bibinfo {title} {{Discovery of ferromagnetism with large magnetic
  anisotropy in ZrMnP and HfMnP}},\ }\href@noop {} {\bibfield  {journal}
  {\bibinfo  {journal} {Appl. Phys. Lett.}\ }\textbf {\bibinfo {volume}
  {109}},\ \bibinfo {pages} {092402} (\bibinfo {year} {2016})}\BibitemShut
  {NoStop}%
\bibitem [{\citenamefont {Klepov}\ \emph {et~al.}(2021)\citenamefont {Klepov},
  \citenamefont {Pace}, \citenamefont {Berseneva}, \citenamefont {Felder},
  \citenamefont {Calder}, \citenamefont {Morrison}, \citenamefont {Zhang},
  \citenamefont {Kirkham}, \citenamefont {Parker},\ and\ \citenamefont
  {Zur~Loye}}]{Klepov2021}%
  \BibitemOpen
  \bibfield  {author} {\bibinfo {author} {\bibfnamefont {V.~V.}\ \bibnamefont
  {Klepov}}, \bibinfo {author} {\bibfnamefont {K.~A.}\ \bibnamefont {Pace}},
  \bibinfo {author} {\bibfnamefont {A.~A.}\ \bibnamefont {Berseneva}}, \bibinfo
  {author} {\bibfnamefont {J.~B.}\ \bibnamefont {Felder}}, \bibinfo {author}
  {\bibfnamefont {S.}~\bibnamefont {Calder}}, \bibinfo {author} {\bibfnamefont
  {G.}~\bibnamefont {Morrison}}, \bibinfo {author} {\bibfnamefont
  {Q.}~\bibnamefont {Zhang}}, \bibinfo {author} {\bibfnamefont {M.~J.}\
  \bibnamefont {Kirkham}}, \bibinfo {author} {\bibfnamefont {D.~S.}\
  \bibnamefont {Parker}},\ and\ \bibinfo {author} {\bibfnamefont {H.-C.}\
  \bibnamefont {Zur~Loye}},\ }\bibfield  {title} {\bibinfo {title} {{Chloride
  Reduction of Mn$^{3+}$ in Mild Hydrothermal Synthesis of a Charge Ordered
  Defect Pyrochlore, CsMn$^{2+}$Mn$^{3+}$F$_6$, a Canted Antiferromagnet with a
  Hard Ferromagnetic Component}},\ }\href@noop {} {\bibfield  {journal}
  {\bibinfo  {journal} {J. Am. Chem. Soc.}\ }\textbf {\bibinfo {volume}
  {143}},\ \bibinfo {pages} {11554} (\bibinfo {year} {2021})}\BibitemShut
  {NoStop}%
\end{thebibliography}

%

\end{document}